\documentstyle[12pt]{article}

\renewcommand{\theequation}{\thesection.\arabic{equation}}

\newcommand \beq{\begin{eqnarray}}
\newcommand \eeq{\end{eqnarray}}
\begin{document}
\input epsf

\def\nbfepsilon{\mbox{\boldmath$\epsilon$}}
\def\nbfgrad{\mbox{\boldmath$\grad$}}
\def\bfgamma{\mbox{\boldmath$\gamma$}}
\def\bfcalA{\mbox{\boldmath${\cal A}$}}
\def\bfcalS{\mbox{\boldmath${\cal S}$}}
\def\bfp{\mbox{\boldmath$p$}}
\def\bfv{\mbox{\boldmath$v$}}
\def\bfj{\mbox{\boldmath$j$}}
\def\bfhp{\mbox{\boldmath$\hat p$}}
\def\bfei{\mbox{\boldmath$e_i$}}
\def\bfe{\mbox{\boldmath$e$}}
\def\bfej{\mbox{\boldmath$e_j$}}
\def\bfk{\mbox{\boldmath$k$}}
\def\bfq{\mbox{\boldmath$q$}}
\def\bfR{\mbox{\boldmath$R$}}
\def\bfC{\mbox{\boldmath$C$}}
\def\bfR{\mbox{\boldmath$R$}}
\def\bfX{\mbox{\boldmath$X$}}
\def\bfx{\mbox{\boldmath$x$}}
\def\bfE{\mbox{\boldmath$E$}}
\def\bfB{\mbox{\boldmath$B$}}
\def\bfy{\mbox{\boldmath$y$}}
\def\bfr{\mbox{\boldmath$r$}}
\def\rmRe{\mbox{\rm$Re$}}
\def\rmIm{\mbox{\rm$ImR$}}

\def\bfgamma{\mbox{\boldmath$\gamma$}}
\def\bfalpha{\mbox{\boldmath$\alpha$}}
\def\bfsigma{\mbox{\boldmath$\sigma$}}
\def\bfalpha{\mbox{\boldmath$\alpha$}}
\def\bfsigma{\mbox{\boldmath$\sigma$}}
\def\bfSigma{\mbox{\boldmath$\Sigma$}}
\def\bfepsilon{\mbox{\boldmath$\epsilon$}}


\def\cad{\hbox{c'est \`a dire}}
\def\Ts{\hbox{temp\'eratures}}
\def\P{\hbox{propri\'et\'e}}
\def\Ps{\hbox{propri\'et\'es}}
\def\E{\hbox{equation}}
\def\LE{\hbox{l'\'equation}}
\def\BTD{\hbox{boucles thermiques dures}}
\def\DV{\hbox{d\'eveloppement}}
\def\DNB{\hbox{d\'eveloppement en nombre de boucles}}
\def\AI{\hbox{amplitudes 1-PI}}
\def\UR{\hbox{ultrarelativiste}}
\def\btd{\hbox{boucle thermique dure}}
\def\em{\hbox{\'electromagn\'etique}}

\def\T{\hbox{temperature}}
\def\bk{\hbox{background}}
\def\P{\mbox{\psi}}
\def\BP{\mbox{\bar\Psi}}
\def\E{\hbox{equation}}
\def\Es{\hbox{equations}}
\def\QGP{\hbox{quark-gluon plasma}}
\def\HTL{\hbox{hard thermal loops}}
\def\htl{\hbox{hard thermal loop}}
\def\se{\hbox{self-energy}}
\def\pt{\hbox{polarization tensor}}
\def\pov{\hbox{point of view}}
\def\pth{\hbox{perturbation theory}}
\def\wr{\hbox{with respect to}}
\def\fn{\hbox{function}}
\def\FN{\hbox{functions}}
\def\BN{\hbox{Bloch-Nordsieck}}
\def\vp{\mbox{$\bf v\cdot p$}}
\def\vq{\mbox{$\bf v\cdot q$}}
\def\vpq{\mbox{$\bf v\cdot(p+ q)$}}
\def\tilA{\mbox{$v\cdot A$}}
\def\tilQ{\mbox{v\cdot q}}
\def\tilQ1{\mbox{$v\cdot q_1$}}
\def\tilQ2{\mbox{$v\cdot q_2$}}
\def\bfp{\mbox{\boldmath$p$}}

\hyphenation{approxima-tions}
\hyphenation{par-ti-cu-le}
\hyphenation{par-ti-cu-les}

\hyphenation{ac-com-pa-gnees}
\hyphenation{cons-tan-te}
\hyphenation{e-lec-tro-ma-gne-ti-que}
\hyphenation{e-lec-tro-ma-gne-ti-ques}
\hyphenation{im-pe-ra-tif}

\newcommand{\theo}{th\'{e}orie\,\,}
\newcommand{\mod}{mod\`ele\,\,}
\newcommand{\mods}{mod\`eles\,\,}
\newcommand{\theos}{th\'{e}ories\,\,}

\def\square{\hbox{{$\sqcup$}\llap{$\sqcap$}}}   
\def\grad{\nabla}                               
\def\del{\partial}                              

\def\frac#1#2{{#1 \over #2}}
\def\smallfrac#1#2{{\scriptstyle {#1 \over #2}}}
\def\half{\ifinner {\scriptstyle {1 \over 2}}
   \else {1 \over 2} \fi}

\def\bra#1{\langle#1\vert}              
\def\ket#1{\vert#1\rangle}              

\def\simge{\mathrel{%
   \rlap{\raise 0.511ex \hbox{$>$}}{\lower 0.511ex \hbox{$\sim$}}}}
\def\simle{\mathrel{
   \rlap{\raise 0.511ex \hbox{$<$}}{\lower 0.511ex \hbox{$\sim$}}}}


\def\parenbar#1{{\null\!                        
   \mathop#1\limits^{\hbox{\fiverm (--)}}       
   \!\null}}                                    
\def\nunubar{\parenbar{\nu}}
\def\ppbar{\parenbar{p}}


\def\buildchar#1#2#3{{\null\!                   
   \mathop#1\limits^{#2}_{#3}                   
   \!\null}}                                    
\def\overcirc#1{\buildchar{#1}{\circ}{}}


\def\slashchar#1{\setbox0=\hbox{$#1$}           
   \dimen0=\wd0                                 
   \setbox1=\hbox{/} \dimen1=\wd1               
   \ifdim\dimen0>\dimen1                        
      \rlap{\hbox to \dimen0{\hfil/\hfil}}      
      #1                                        
   \else                                        
      \rlap{\hbox to \dimen1{\hfil$#1$\hfil}}   
      /                                         
   \fi}                                         %


\def\subrightarrow#1{
  \setbox0=\hbox{
    $\displaystyle\mathop{}
    \limits_{#1}$}
  \dimen0=\wd0
  \advance \dimen0 by .5em
  \mathrel{
    \mathop{\hbox to \dimen0{\rightarrowfill}}
       \limits_{#1}}}                           

\def\real{\mathop{\rm Re}\nolimits}     
\def\imag{\mathop{\rm Im}\nolimits}     

\def\tr{\mathop{\rm tr}\nolimits}       
\def\Tr{\mathop{\rm Tr}\nolimits}       
\def\Det{\mathop{\rm Det}\nolimits}     

\def\mod{\mathop{\rm mod}\nolimits}     
\def\wrt{\mathop{\rm wrt}\nolimits}     


\def\TeV{{\rm TeV}}                     
\def\GeV{{\rm GeV}}                     
\def\MeV{{\rm MeV}}                     
\def\KeV{{\rm KeV}}                     
\def\eV{{\rm eV}}                       

\def\mb{{\rm mb}}                       
\def\mub{\hbox{$\mu$b}}                 
\def\nb{{\rm nb}}                       
\def\pb{{\rm pb}}                       

%
\def\journal#1#2#3#4{\ {#1}{\bf #2} ({#3})\  {#4}}

\def\AdvPhys{\journal{Adv.\ Phys.}}
\def\AnnPhys{\journal{Ann.\ Phys.}}
\def\EurophysLett{\journal{Europhys.\ Lett.}}
\def\JApplPhys{\journal{J.\ Appl.\ Phys.}}
\def\JMathPhys{\journal{J.\ Math.\ Phys.}}
\def\LettNuovoCimento{\journal{Lett.\ Nuovo Cimento}}
\def\Nature{\journal{Nature}}
\def\NPA{\journal{Nucl.\ Phys.\ {\bf A}}}
\def\NPB{\journal{Nucl.\ Phys.\ {\bf B}}}
\def\NuovoCimento{\journal{Nuovo Cimento}}
\def\Physica{\journal{Physica}}
\def\PLA{\journal{Phys.\ Lett.\ {\bf A}}}
\def\PLB{\journal{Phys.\ Lett.\ {\bf B}}}
\def\PR{\journal{Phys.\ Rev.}}
\def\PRC{\journal{Phys.\ Rev.\ {\bf C}}}
\def\PRD{\journal{Phys.\ Rev.\ {\bf D}}}
\def\PRB{\journal{Phys.\ Rev.\ {\bf B}}}
\def\PRL{\journal{Phys.\ Rev.\ Lett.}}
\def\PhysRept{\journal{Phys.\ Repts.}}
\def\ProcNatlAcadSci{\journal{Proc.\ Natl.\ Acad.\ Sci.}}
\def\ProcRoySoc{\journal{Proc.\ Roy.\ Soc.\ London Ser.\ A}}
\def\RevModPhys{\journal{Rev.\ Mod.\ Phys. }}
\def\Science{\journal{Science}}
\def\SovPhysJETP{\journal{Sov.\ Phys.\ JETP }}
\def\SovPhysJETPLett{\journal{Sov.\ Phys.\ JETP Lett. }}
\def\SovJNuclPhys{\journal{Sov.\ J.\ Nucl.\ Phys. }}
\def\SovPhysDoklady{\journal{Sov.\ Phys.\ Doklady}}
\def\ZPhys{\journal{Z.\ Phys. }}
\def\ZPhysA{\journal{Z.\ Phys.\ A}}
\def\ZPhysB{\journal{Z.\ Phys.\ B}}
\def\ZPhysC{\journal{Z.\ Phys.\ C}}



\begin{titlepage}
\begin{flushright} {Saclay-T96/085}
\end{flushright}

\vspace*{0.2cm}
\begin{center}
\baselineskip=20pt {\Large LIFETIMES OF  QUASIPARTICLES\\

\bigskip

AND COLLECTIVE EXCITATIONS

\bigskip

 IN HOT QED PLASMAS}

 \vskip0.5cm Jean-Paul
BLAIZOT\footnote{CNRS}
  and Edmond IANCU\footnote{CNRS}
\\ {\it Service de Physique Th\'eorique\footnote{Laboratoire
 de la Direction des Sciences de
la Mati\`ere du Commissariat \`a l'Energie Atomique},
CE-Saclay \\ 91191 Gif-sur-Yvette, France}

\end{center}

\vskip 1cm
\vskip 1cm
\begin{abstract}
The perturbative calculation of the lifetime of fermion excitations 
in a QED plasma at high temperature is plagued with infrared
divergences which are not eliminated by the screening corrections.
  The physical processes responsible for these
divergences are the collisions involving the exchange of
longwavelength, quasistatic, magnetic photons, which
are not screened by plasma effects.
The leading divergences can be resummed
 in a non-perturbative treatement based on
 a generalization of the Bloch-Nordsieck model at finite
temperature. The resulting expression of the fermion
 propagator is free of infrared problems, and exhibits
a {\it non-exponential} damping at large times:
$S_R(t)\sim
\exp\{-\alpha T \, t\, \ln\omega_pt\}$, where $\omega_p=eT/3$ is the plasma frequency
and $\alpha=e^2/4\pi$.
\end{abstract}

\vskip 2.6cm

\begin{flushleft}
Submitted to Physical Review {\bf D}
\end{flushleft}

\end{titlepage}


\baselineskip=20pt

\section{Introduction}
\setcounter{equation}{0}

The study of the elementary excitations of ultrarelativistic  plasmas,
such as  the quark-gluon plasma,  has received much
attention in the recent past [1---11]
(See also \cite{BIO96,MLB96} for recent reviews and more references.)
The physical picture which emerges is that of a system with
 two types of degrees of freedom:
{\it i}) the plasma quasiparticles,
whose energy is of the order of the temperature $T$;
{\it ii}) the collective excitations, whose typical energy
is $gT$, where $g$ is the gauge coupling,
assumed to be small: $g\ll 1$ (in QED, $g=e$ is the electric charge).
For this picture to make sense, however, it is important that the
 lifetime of the excitations be large compared to the
typical period of the modes. 

Information about the lifetime is obtained from the
retarded propagator. A usual expectation is that
 $S_R(t,{\bf p})$ decays {\it exponentially} in time,
 $S_R(t,{\bf p})\,\sim\,{\rm e}^{-i E(p)t} {\rm e}^{ -\gamma({p}) t}$,
so that  $|S_R(t,{\bf p})|^2\,\sim\,{\rm e}^{ -\Gamma({p}) t}$ 
with $\Gamma(p)=2\gamma(p)$, which identifies the lifetime
of the single particle excitation as $\tau(p) = 1/\Gamma(p)$.
The exponential decay may then be associated to a pole
of the Fourier transform  $S_R(\omega,{\bf p})$,
located at $\omega = E(p)-i\gamma(p)$.
The quasiparticles are well defined
if their lifetime  $\tau$ is much larger than the period $\sim 1/E$
of the field oscillations, that is, if the damping rate
$\gamma$  is small compared to the energy $E$. If this is the case,
 the respective damping rates
can be computed from the imaginary part of the on-shell
self-energy, $\Sigma(\omega=E(p), {\bf p})$. Such calculations suggest that 
$\gamma\sim g^2T$ \cite{Pisarski89,BP90}
for both the single-particle and the collective excitations.
In the weak coupling regime $g\ll 1$,
 this is indeed small compared to the corresponding
energies (of order $T$ and $gT$, respectively),
suggesting that the  quasiparticles are well defined, and the
collective modes are weakly damped. However, the computation of 
$\gamma$ in perturbation theory 
is plagued with infrared divergences, which casts doubt on the
validity of these statements \cite{Pisarski89}, [15---26]

The first attempts to calculate the damping rates
were made in the early 80's. It was then found that,
to one-loop order, the damping rate of the soft
collective excitations in the hot QCD plasma was gauge-dependent,
and could turn out negative in some gauges (see Ref. \cite{Pisarski91}
for a survey of this problem). Decisive progress on this 
problem was made by Pisarski \cite{Pisarski89} and Braaten and
Pisarski who identified the resummation needed to obtain the
screening corrections in a gauge-invariant way \cite{BP90}
(the resummation of the so called ``hard thermal loops'' (HTL)).
Such screening corrections are sufficient to
make finite the transport cross-sections\cite{Baym90,BThoma91},
and also the damping rates of excitations 
with zero momentum\cite{BP90,KKM}. At the same time, however,
it has been remarked\cite{Pisarski89} that the HTL resummation
 is not sufficient to render finite
 the damping rates of excitations with non vanishing momenta.
The remaining infrared divergences are due to collisions involving the
exchange of longwavelength, quasistatic, magnetic photons (or gluons),
which are not screened in the hard thermal loop approximation.
Such divergences affect the computation of the damping rates
of {\it charged} excitations, in both abelian and non-abelian
gauge theories. Thus, in the lowest order calculations
of Refs. \cite{Pisarski89}, [15---26], one
meets the same logarithmic divergence for electrons in QED,
for charged scalars in SQED, and for quarks and gluons in QCD.
(There is no such problem for the photon damping rate,
which is IR finite and of order $g^4T$ \cite{Thoma94}, since 
photons do not couple directly to gluons or to themselves.)
Furthermore, the problem appears for both soft ($p \sim gT$) and hard 
 ($p \sim T$) quasiparticles. In QCD this problem is generally
 avoided by the {\it ad-hoc} introduction of an IR cut-off
(``magnetic screening mass'') $\sim g^2T$, which is
expected to appear dynamically from gluon
self-interactions \cite{Linde80}.
In QED, on the other hand, it is known that no magnetic
screening can occur\cite{Fradkin65},
so that the solution of the problem must lie somewhere else.

In order to make the damping rate $\gamma$ finite,
 Lebedev and Smilga proposed
a  self-consistent computation of  the damping rate $\gamma$\cite{Lebedev90}, 
by including $\gamma$  also in internal propagators.
However, the resulting self-energy
is not analytic near the complex mass-shell, and the logarithmic
divergence actually reappears when the discontinuity of the self-energy is evaluated 
at $\omega= E - i\gamma$  \cite{Baier92,Pisarski93}. More
thorough resummations of the fermion line led to the conclusion
that the full fermion propagator has actually  no quasiparticle
pole in the complex energy plane\cite{Smilga93,Pilon93}. These analyses left
unanswered, however, the question of the large time
behavior of the retarded propagator.
As we have shown in a previous letter\cite{prl}, the answer to this question
 requires resummations for both the fermion propagator
and the photon-electron vertex function.
Such resummations modify the analytic structure of
the  retarded propagator: indeed, as we shall see, they make it analytic
in the vicinity of the mass-shell.

The need for a nonperturbative analysis follows from the fact that 
infrared divergences  occur in {\it all} orders of perturbation theory.
The {\it leading} divergences arise,
in all orders, from the same kinematical regime as in the
one loop calculation, namely from the exchange of soft {\it quasistatic}
magnetic photons. In the imaginary time formalism, 
these divergences are concentrated in diagrams in which the
 photon lines carry zero Matsubara frequency (to be referred as
{\it static modes} in what follows). 
In this sense, they appear as the divergences of
 an effective three-dimensional gauge theory, which is
intrinsically non-perturbative. Still, this effective
``dimensional reduction'' brings in simplifications which can be
exploited to arrive at an explicit solution of the problem.

We concentrate in this paper on the damping rate of fermionic
excitations in hot QED plasmas. Our analysis
is based on the Bloch-Nordsieck (or eikonal) approximation\cite{BN37}.
At zero temperature, this approximation provides an all-order solution
to the infrared catastrophe, and  correctly  describes 
the mass-shell structure of
the 4-dimensional fermion propagator\cite{Bogoliubov}. 
At finite temperature, the Bloch-Nordsieck approximation
has been previously used, by Weldon, to verify the cancellation
 of the infrared divergences
in the production rate for soft real photons \cite{Weldon94}.
Let us also mention that  an attempt
to solve the IR problem of the damping rate,
using the BN approximation in the same spirit as in the present
paper, has been reported in Ref.  \cite{Takashiba95}. However,
although the final result obtained in \cite{Takashiba95}
is similar to ours, the derivations there are plagued
with several inconsistencies, some of which are pointed
out in \cite{prl}.

In this paper, we shall consider (in section 3) a different  generalization
of the Bloch-Nordsieck (BN) model at finite temperature,
which is better suited to study the infrared structure
of the fermion propagator. Our approach is a natural
extension of the method used in
Ref. \cite{Bogoliubov} in  ${\rm QED}_{3+1}$ at zero temperature. 
However, the resulting imaginary-time BN propagator does not exponentiate
in an obvious way, and thus cannot be written
in closed form, in contrast to the usual, zero-temperature propagator.
Still,  we can obtain an explicit solution
 once we restrict ourselves to the static Matsubara photon
modes. We thus get the retarded propagator $S_R(t,{\bf p})$, 
and study its large time behavior (section 4).
 The final result is that, for times $t\gg 1/gT$,
the propagator does not show the usual exponential decay alluded
to before, but the more complicated behavior
$S_R(t,{\bf p})\,\sim\,{\rm e}^{-iE(p) t} {\rm e}^{-\alpha T \, t\,
\ln\omega_pt}$, where $\omega_p\sim gT$ is the plasma frequency,
and $E(p)\simeq p \sim T$ is the average energy of the hard fermion.
This corresponds to a typical lifetime $\tau^{-1}\sim g^2T\ln (1/g)$,
which is similar to the one provided by the perturbation
theory with an IR cut-off of the order $g^2T$. 
Since, as $t\to \infty$,
$S_R(t)$ is decreasing faster than any exponential, the
 Fourier transform of $S_R(t,{\bf p})$,
 $S_R(\omega,{\bf p})$, is an entire function in the complex energy plane. 
The existence of the quasiparticle is therefore not signaled
by the presence of a pole of  $S_R(\omega)$ in the complex
energy plane. However, the  associated
spectral density has the shape of a
{\it resonance}  strongly peaked around $\omega = E(p)$,
 with a typical width of the order $1/\tau\sim g^2T \ln(1/g)$.
With minor modifications, the above conclusions also apply
for the {\it soft} (collective) excitations, with momenta $p\sim gT$,
whose lifetimes are found to depend on the group velocities $|v_\pm|< 1$
(section 5).

At this stage it is useful to specify the notations and the conventions
to be used throughout. The analytic propagator is defined
in the complex energy plane by the spectral representation
\beq\label{Sspec0}
S(\omega, {\bf p})&=&\int_{-\infty}^{+\infty}\frac{{\rm d}p^0}{2\pi}\,
\frac{\rho_{\rm f}(p^0, {\bf p})}{p^0-\omega}\,.\eeq
 The  Matsubara propagator is obtained
from eq.~(\ref{Sspec0}) by setting $\omega= i\omega_n$,
with $\omega_n=(2n+1)\pi T$ and integer $n$. 
 At tree level, $\rho_{\rm f}(p^0, {\bf p})
= {\slashchar p} \rho_0(p^0, p)$ where ${\slashchar p}= p^\mu \gamma_\mu$,
$\epsilon_p \equiv |{\bf p}| =p$, and 
 \beq\label{rho0}\rho_0(p^0,p)\,=\,
\frac{\pi}{\epsilon_p}\,\Bigl(\delta(p^0-\epsilon_p)\,-\,
\delta(p^0+\epsilon_p)\Bigr), \eeq 
so that
\beq\label{S0}
S_0(\omega, {\bf p})=-\frac {\omega\gamma^0 -{\bf p}\cdot {\bfgamma}}
{\omega^2-p^2}=\,
\frac{-1}{\omega- p}\, h_+(\hat {\bf p})
+ \frac{-1}{\omega+ p}\, h_-(\hat {\bf p}),\eeq
where  $h_\pm(\hat {\bf p}) = (\gamma^0\mp \hat {\bf p}\cdot \bfgamma)/2$,
with ${\hat{\bf  p}}\equiv {\bf  p}/p$.

The full fermion propagator is given by the Dyson-Schwinger
equation 
\beq\label{SD}
S^{-1}(\omega, {\bf p})\,=\,S^{-1}_0(\omega, {\bf p})
\,+\,\Sigma(\omega, {\bf p}).\eeq
The most general form of the self-energy $\Sigma$ which is compatible
with the rotational and chiral symmetries is
 \beq\label{Sigma1}
\Sigma(\omega, {\bf  p})\,=\,a(\omega, p)\,\gamma^0\,+\,b(\omega, p)
{\hat{\bf  p}}\cdot{\bfgamma}\equiv 
h_-(\hat {\bf p})\Sigma_+(\omega,p)\,-\,
h_+(\hat {\bf p})\Sigma_-(\omega,p),\eeq
where
\beq\label{PIpm}
\Sigma_\pm(\omega,p) &=&\pm \,\frac{1}{2}\,{\rm tr}\,\Bigl(
h_\pm(\hat {\bf p}) \Sigma(\omega, {\bf p})\Bigr).\eeq
Using this decomposition of $\Sigma$ onto $h_\pm$, and the analogous
one for $S_0$, eq.~(\ref{S0}), one can easily invert eq.~(\ref{SD}) to get
the full propagator:
\beq\label{Sfull}
S(\omega, {\bf p})=\Delta_+(\omega,p) h_+(\hat {\bf p})
+\Delta_-(\omega,p) h_-(\hat {\bf p}),\eeq
where
\beq\label{DPM}
\Delta_\pm(\omega,p) &=&\frac{-1}{\omega\mp (p + \Sigma_\pm(\omega,p))}\,.\eeq

The retarded propagator is obtained as the boundary value
of the analytic propagator  (\ref{Sspec0}) when $\omega$
approaches the real axis from above, i.e.,
$S_R(\omega, {\bf p})=S(\omega +i\eta, {\bf p})$, where
$\omega$ is real and $\eta\to 0_+$.
In the time representation, 
\beq\label{GRTrho}
S_R(t, {\bf p})\,=\,
\int_{-\infty}^\infty {{\rm d}\omega \over 2\pi}\,{\rm e}^{-i\omega t}
S_R(\omega, {\bf p})\,=\,i\theta(t)
\int_{-\infty}^\infty {{\rm d}\omega \over 2\pi}\,{\rm e}^{-i\omega t}
\rho_{\rm f}(\omega,{\bf p}).\eeq
The large time behavior of $S_R(t,{\bf p})$ is determined by
the analytic structure of $S_R(\omega, {\bf p})$ when continued to 
complex values of $\omega$.
In the upper half plane, $S_R(\omega)$ coincides with the analytic
propagator (\ref{Sspec0}). In  the lower half plane,   $S_R(\omega)$ is
defined by continuation across the real axis, and it may have there
singularities. The large time behavior
of  $S_R(t)$ is controlled in most cases by the  singularity of  $S_R(\omega)$
which lies closest to the real axis.
If this is located at $\omega = E(p)-i\gamma(p)$, then 
 $S_R(t,{\bf p})\,\sim\,f(t,{\bf p})\,
{\rm e}^{-i E(p)t} {\rm e}^{ -\gamma({p}) t}$,
where the prefactor $f(t,{\bf p})$ is slowly varying,
and depends on the specific nature of the singularity.
This conventional picture breaks down in gauge theories
since, as we shall discuss in the next section,
the perturbative estimate of $\gamma$ turns out to be IR divergent.
The resummation of the leading infrared divergences,
carried out in section 3, produces a propagator
which has no  singularity in the complex $\omega$ plane.
We shall then find it convenient to calculate $S_R(t)$ directly,
rather than  from the Fourier transform (\ref{GRTrho}).

\setcounter{equation}{0}

\section{The one-loop damping rate for the hard fermion}

In this section, we review the perturbative calculations of the damping rate
for a hard fermion, with momentum $p\sim T$ [15---26]
focussing on the infrared divergences which arise in such calculations.
We assume here, as customary, that the dominant singularity
of the retarded propagator
is a simple pole whose location  goes back into the tree-level
pole at $\omega=p$ when $g\to 0$.


\subsection{Physical interpretation of the damping}

\begin{figure}
\protect \epsfysize=2.cm{\centerline{\epsfbox{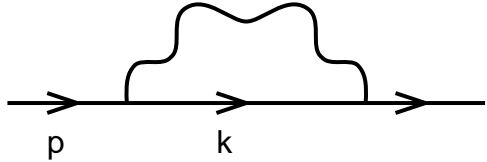}}}
	 \caption{The one-loop fermion self-energy}
\label{figfermion}
\end{figure}
To leading order in $g$, the self-energy is given by the
one-loop diagram in Fig.~\ref{figfermion}. This gives no contribution
to the damping rate $\gamma$. Indeed, when evaluated on the free
 mass-shell, i.e. at $\omega = p$, the imaginary part of the one-loop
self-energy vanishes because of kinematics. (At finite
temperature this argument involves 
subtleties which are discussed in Appendix B.)

\begin{figure}
\protect \epsfxsize=10.cm{\centerline{\epsfbox{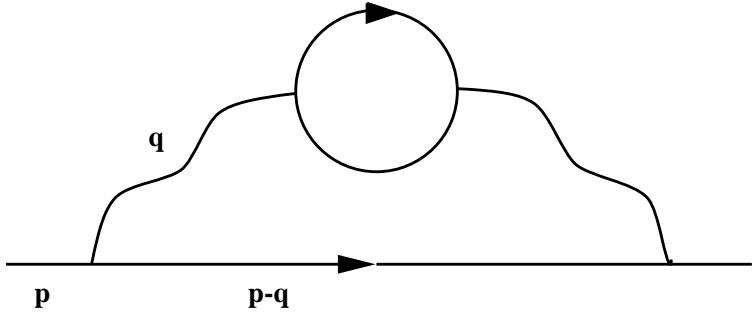}}}
	 \caption{Two-loop diagram contributing to the fermion self-energy}
\label{fermion2l}
\end{figure}

The leading contribution to $\gamma$ comes therefore
from the two-loop diagram in Fig.~\ref{fermion2l}, and turns out
to be quadratically infrared divergent
(see, e.g., Refs. \cite{Baym90,BThoma91,Altherr93,MLB96}).

The on-shell imaginary part is obtained by cutting the
diagram in Fig.~\ref{fermion2l} through the internal fermion loop and the
lower fermion propagator. Physically, this imaginary part accounts
for the scattering of the incoming electron (with four momentum
$p^\mu=(\epsilon_p, {\bf p})$ and $\epsilon_p =p$)
off a thermal fermion (electron or positron), calculated
in the Born approximation (see Fig. 3). The total interaction rate
is given by
\beq\label{gammaB}
\Gamma(p)&=&\frac{1}{2\epsilon}
\int {\rm d}\tilde p_1 \,{\rm d}\tilde p_2\,{\rm d}\tilde p_3\,
(2\pi)^4\delta^{(4)}(p+p_1-p_2-p_3)\nonumber\\
&{}&\,\,\,\,\Bigl\{n_1(1-n_2)(1-n_3)+(1-n_1)n_2 n_3\Bigr\}\,
{|{\cal M}|^2},\eeq
and coincides with twice the damping rate $\gamma(p)$,
as computed from the two-loop self-energy in Fig. 2: $\Gamma(p)=2\gamma(p)$.
This identity extends to finite temperature the usual
physical interpretation of the self-energy discontinuity
in terms of cross-sections for physical processes, and  
 can be verified through an explicit calculation\cite{MLB96}
(see also below).
\begin{figure}
\protect \epsfxsize=7.cm{\centerline{\epsfbox{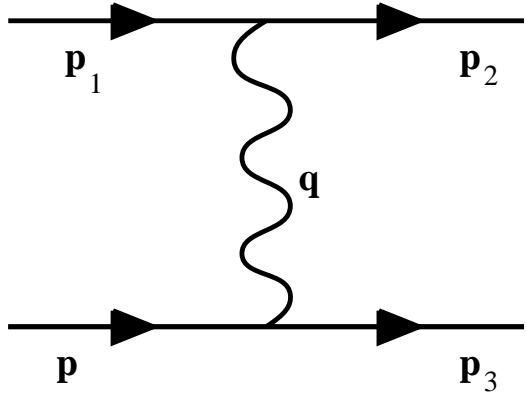}}}
	 \caption{Fermion-fermion elastic scattering in the Born approximation}
\label{Born}
\end{figure}
The notations in eq.~(\ref{gammaB}) are as follows:
 all the particles are on the mass-shell
(i.e., $\epsilon=p$ and $\epsilon_i=p_i$ for $i=1,\,2,\,3$),
and we have denoted $\int {\rm d}\tilde p_i \equiv
 \int ({\rm d}^3p_i/(2\pi)^3\,2\epsilon_i)$.
The factors $n_i=n(\epsilon_i)$ are the thermal occupation numbers 
for fermions ($n(\epsilon)=1/({\rm e}^{\beta \epsilon}+1)$).
Note that, for fermions, the rates of the direct 
and of the reverse  processes
have to be {\it added}  to give the total
depopulation of the fermion state with momentum $p^\mu$
\cite{KB62}. Finally,  ${|{\cal M}|^2}$ is the 
scattering matrix element squared, 
 averaged over the spin $s$ of the incoming electron,
and summed over the  spins $s_1$,  $s_2$ and  $s_3$
of the other three particles. In the Born
 approximation,  ${|{\cal M}|^2}$ is independent
of the temperature and involves the 
 propagator $D_{\mu\nu}(q)$ of the exchanged photon
(with $q=p-p_3=p_2-p_1$). Specifically\cite{BD62},
\beq\label{Mborn}
|{\cal M}|^2= 16g^4  D_{\mu\nu}(q)D^*_{\rho\lambda}(q)
\Bigl\{p^\mu p_3^\rho + p_3^\mu p^\rho - g^{\mu\rho}(p\cdot p_3)\Bigr\}
 \Bigl\{p_1^\nu p_2^\lambda + p_2^\nu p_1^\lambda - g^{\nu\lambda}
(p_1\cdot p_2)\Bigr\}\,.\eeq
 We shall use below the Coulomb gauge where the only non-trivial
components of $D_{\mu\nu}(q)$
 are the electric (or longitudinal)
one $D_{00}(q)\equiv \Delta_l(q)$, and the magnetic (or transverse) one
$D_{ij}(q)=(\delta_{ij}-\hat q_i\hat q_j)\Delta_t(q)$.

Since the interaction rate (\ref{gammaB}) is dominated by 
soft momentum transfers $q\ll T$, while the external
 momenta  are typically of the order of $T$, we can
 simplify the matrix element $|{\cal M}|^2$
by setting ${\bf p}\simeq {\bf p_3}$ and ${\bf p_1}\simeq {\bf p_2}$
in eq.~(\ref{Mborn}), and obtain
\beq |{\cal M}|^2\simeq 64 g^4 p^2 p_1^2\,\Big|\Delta_l(q)
+ ({\bf v \times \hat q})\cdot ({\bf v_1 \times \hat q})
 \Delta_t(q)\Big|^2,\eeq
 with ${\bf v \equiv \hat p}$ and ${\bf v_1 \equiv \hat p_1}$.
Furthermore, we use energy conservation to write
 $q_0=\epsilon-\epsilon_3=\epsilon_2-\epsilon_1$,
 that is,
 $$q_0=p-|{\bf p-q}|=|{\bf p_1+q}|-p_1\,,$$
which, for $q\ll T$, becomes 
\beq\label{encons}
q_0\simeq {\bf v\cdot q} \simeq {\bf v_1\cdot q}\,.\eeq
The statistical factors in eq.~(\ref{gammaB}) satisfy
 the following identity:
\beq\label{FDBE}
n_1(1-n_2)(1-n_3)+(1-n_1)n_2 n_3\,=\,(n_1-n_2)(1 + N(q_0)-n_3)\,,\eeq
which features $N(q_0)$, the Bose-Einstein thermal factor for the virtual photon.
Since $\epsilon_2=\epsilon_1+q_0$, and
  $q_0\ll p_1\sim T$, 
\beq
(n_1-n_2)(1 + N(q_0)-n_3)\,\simeq\,-\frac{{\rm d}n}{{\rm d} p_1}\,q_0 N(q_0)\,
\simeq \,-T \frac{{\rm d}n}{{\rm d}p_1}\,,\eeq
where we have used the fact that, at small $q_0\ll T$,
\beq
1+N(q_0)-n_3\,\simeq\, N(q_0)\,\simeq\, T/q_0\,.\eeq
Finally, we use eq.~(\ref{encons}) to rewrite
the integrations over ${\bf p_2}$ and ${\bf p_3}$
 as follows
\beq
\int \frac{{\rm d}^3p_2}{(2\pi)^3}\int \frac{{\rm d}^3p_3}{(2\pi)^3}
\,\,(2\pi)^4\delta^{(4)}(p+p_1-p_2-p_3)\qquad\nonumber\\
=\,\,\int \frac{{\rm d}^3q}{(2\pi)^3}
 \int_{-\infty}^\infty \frac{{\rm d}q_0}{2\pi}\,
{2\pi}\delta(q_0 -  {\bf v\cdot q})\,{2\pi}\delta(q_0 -  {\bf v_1\cdot q}),\eeq
so that we may use  ${\bf p_1}$, ${\bf q}$ and  $q_0$
 as independent  integration variables
in eq.~(\ref{gammaB}):
\beq\label{G20}
\Gamma(p)&\simeq& 16\pi^2 g^4 T\int\frac{{\rm d}^3p_1}{(2\pi)^3}\left
(- \frac{{\rm d}n}{{\rm d}p_1}\right)\,
\int \frac{{\rm d}^3q}{(2\pi)^3}
\int_{-\infty}^\infty \frac{{\rm d}q_0}{2\pi}\,
\nonumber\\ &{}&
\delta(q_0 -  {\bf v\cdot q})\,\delta(q_0 -  {\bf v_1\cdot q})\,
\Big|\Delta_l(q)
+ ({\bf v \times \hat q})\cdot ({\bf v_1 \times \hat q})
 \Delta_t(q)\Big|^2.\eeq
We perform the angular integrations over ${\bf v_1 \equiv \hat p_1}$
and ${\bf \hat q}$   by using the delta functions,
while the radial integration over $p_1$ gives
\beq \int {\rm d}p_1 p_1^2 \left
(- \frac{{\rm d}n}{{\rm d}p_1}\right)\,=\,\frac{\pi^2 T^2}{6}\,.\eeq
We  obtain finally
\beq\label{G2L}
\Gamma \simeq\, \frac{g^4 T^3}{6}\,
\int_{0}^{q^*}{\rm d}q  \int_{-q}^q\frac{{\rm d}q_0}{2\pi}
\left\{ |\Delta_l(q_0,q)|^2\,+\,\frac{1}{2}\left(1-\frac{q_0^2}{q^2}\right)^2
|\Delta_t(q_0,q)|^2
\right\}\,,\eeq
where the upper cut-off $q^*$ distinguishes between
soft and hard momenta: $gT\ll q^* \ll T$.
Since the $q$-integral is dominated by IR momenta, its leading
order value is actually independent of $q^*$.

The two terms within the parentheses in eq.~(\ref{G2L})
  correspond to the exchange of an electric and of
a magnetic photon respectively.
For a bare photon, we have $|\Delta_l(q_0,q)|^2= 1/q^4$ and
$|\Delta_t(q_0,q)|^2= 1/(q_0^2-q^2)^2$, so that
 the $q$-integral in eq.~(\ref{G2L}) shows a quadratic IR divergence:
\beq\label{G2L0}
\Gamma\simeq  \frac{g^4T^3}{4\pi} \,
\int_{0}^{q^*}\frac{{\rm d}q}{q^3}\,.\eeq
This divergence reflects the singular behaviour
of the Rutherford cross-section for forward scattering \cite{BD62}.

As well known, however, the quadratic divergence is removed by the
screening corrections contained in the photon polarization tensor.
 These modify the electric and magnetic propagators as follows
 \beq\label{effd}
{}^*\Delta_l(q_0,q)\,=\,\frac{- 1}{q^2- \delta\Pi_l(q_0,q)},\qquad
{}^*\Delta_t(q_0,q)\,=\,\frac{-1}{q_0^2-q^2 -\delta\Pi_t(q_0,q)},\eeq
where $\delta\Pi_l$ and $\delta\Pi_t$ are the respective pieces
of the photon polarisation tensor
 (in the hard thermal loop approximation\cite{Klimov81,Weldon82}).
We shall see below that the leading IR contribution
comes from the domain $q_0\ll q \ll T$, where we can
 use the approximate expressions
\beq\label{pltstatic}
\delta\Pi_l(q_0\ll q) \simeq  3{\omega_p^2}\,\equiv m_D^2,\qquad
\delta\Pi_t(q_0\ll q) \simeq \,-i\,\frac{3\pi}{4}\,{\omega_p^2}\,\frac{q_0}{q}\,
.\eeq 
We see that screening occurs in different ways
in the electric and the magnetic sectors.
In the electric sector, the familiar static Debye screening provides
an IR cut-off $m_{D}\sim gT$. Accordingly,
 the electric contribution to $\Gamma$ is finite,
and of the order  $\Gamma_l \sim g^4 T^3/m_{D}^2
\sim g^2 T$. Its exact value can be computed by numerical
integration\cite{Pisarski93}. In the magnetic sector,
  screening occurs only for nonzero frequency $q_0$ \cite{Weldon82,Baym90}.
This comes from the imaginary part of the polarisation tensor,
and can be associated to the Landau damping of space-like photons ($q_0^2<q^2$).
This  ``dynamical screening'' is not sufficient to completely
remove the IR divergence of $\Gamma_t\,$:
\beq\label{G2LR}
\Gamma_t &\simeq& \frac{g^4 T^3}{12}\,
\int_{0}^{q^*}{\rm d}q  \int_{-q}^q\frac{{\rm d}q_0}{2\pi}
\,\frac{1}{q^4 + (3\pi \omega_p^2 q_0/4q)^2} \nonumber\\
&=&\frac{g^2T}{\pi^2}\,\int_{0}^{q^*}\frac{{\rm d}q}{q}\,
\arctan\left(\frac{3\pi\omega_p^2}{4q^2}\right)\simeq\,
\frac{g^2T}{2\pi}\,\int_{0}^{\omega_p}\frac{{\rm d}q}{q}\,.\eeq
In writing the last equality, we payed attention only
to the dominant, logarithmically divergent, contribution.
To isolate it, we have written
$$ \arctan\left(\frac{3\pi\omega_p^2}{4q^2}\right)\simeq\,\frac{\pi}{2}\,,$$
as appropriate for $q\ll \omega_p$, and we have introduced the upper
 cut-off $\omega_p\sim gT$  to
 approximately account for the correct UV behaviour
of the integrand: namely, as $q\gg \omega_p$,
the integrand is decreasing like $\omega_p^2/q^3$, so that the 
$q$-integral is indeed cut-off at $q\sim \omega_p$.

The remaining IR divergence in eq.~(\ref{G2LR}) is due to collisions involving the
exchange of very soft ($|{\bf q\to 0}|$),
 {\it quasistatic} ($q_0\to 0$) magnetic photons,
which are not screened by plasma effects.
To see that, note that the IR contribution to
 $\Gamma_t$ comes from momenta $q\ll gT$,
where $|\Delta_t(q_0,q)|^2$ is almost a delta function of $q_0$:
\beq \label{singDT}
|\Delta_t(q_0,q)|^2\,\simeq\,
\frac{1} {q^4 + (3\pi \omega_p^2 q_0/4q)^2}\,
\longrightarrow_{q\to 0}\,\frac{4}{3 q \omega_p^2}\,\delta(q_0)\,.\eeq
This is so because,
as $q_0\to 0$, the imaginary part of the polarisation
tensor vanishes {\it linearly}
(see the second equation (\ref{pltstatic})), 
a property which can be related to the behaviour of the
phase space for the Landau damping processes.
Since energy conservation requires $q_0=q\cos\theta$, 
where $\theta$ is the angle
between the momentum of the virtual photon (${\bf q}$) and that
of the incoming fermion (${\bf p}$), 
the magnetic photons which are responsible for the singularity
are emitted, or absorbed, at nearly 90 degrees.

To conclude this subsection, we note that, if we temporarily
 leave aside the logarithmic divergence, then both
the electric and the magnetic damping rates are of order $g^2T$,
rather than $g^4T$ as one would naively expect
by looking at the diagrams in Figs. 2 and 3. This situation
has been sometimes referred as {\it anomalous damping} \cite{Lebedev90},
and is a consequence of the strong sensitivity of the  scattering
cross section  to the IR behavior of the photon propagator.
By comparison,  the other two-body collisions leading to
the damping of the fermion, namely the Compton
scattering and the annihilation process, are less IR singular
--- as they involve the exchange of
a virtual {\it fermion} --- and only contribute
at order $g^4 T$.

\subsection{Resummed one-loop self-energy}

While the above calculation of the interaction rate
in the Born approximation is physically transparent, for the
subsequent developments in this paper it is
more convenient to obtain $\gamma$ from
the imaginary part of the self-energy.
To lowest order, we can write
  $\gamma(p)= - {\rm Im}\, \Sigma_+(p,p)$,
with $ \Sigma_+(\omega,p)$  defined as in eq.~(\ref{PIpm}) in terms
of the {\it resummed} one-loop self-energy.
\begin{figure}
\protect \epsfxsize=11.cm{\centerline{\epsfbox{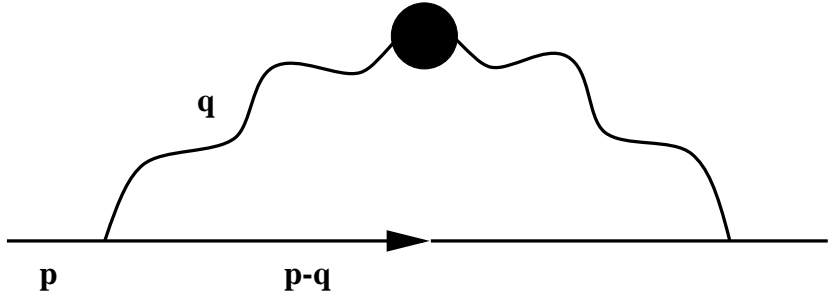}}}
	 \caption{The resummed one-loop self-energy}
\label{resummedfermion}
\end{figure}
 The corresponding diagram is displayed in Fig.~\ref{resummedfermion}:
the blob on the photon line  in this figure
 denotes the  effective  photon propagator of eq.~(\ref{effd}).

To evaluate the one-loop diagram in Fig.~\ref{resummedfermion}, we use the 
imaginary time formalism  and write
\beq\label{Sigeff}
\Sigma(p)&=&-\,g^2 T\sum_{q^0= i\omega_m}
 \int \frac{{\rm d}^3q}{(2\pi)^3}\,
\,\gamma_\mu\,S_0(p-q)\,\gamma_\nu
\,{}^*D^{\mu\nu}(q)\,.\eeq
In this equation, all the energy variables
are purely imaginary and discrete to start with; namely,
 $p^0= i\omega_n= i(2n+1)\pi T$ for the external fermion line,
and  $q^0= i\omega_m= i2\pi mT$ for the internal photon line,
 with integers $n$ and $m$.
Furthermore,    ${\bf k=p-q}$, $S_0(p-q)$ is the free fermion propagator,
eq.~(\ref{S0}), and ${}^*D^{\mu\nu}(q)$ is the resummed photon propagator.
 We shall perform our computations in the  Coulomb gauge
(the one-loop damping rate is gauge 
independent\cite{BP90,KKR90,Rebhan93}; see also Appendix B).

 The continuation of $\Sigma(p)$ to real external
energy can be done only {\it after} performing
the Matsubara sum over $q_0=i\omega_m$, and consists in
 simply replacing (for retarded boundary conditions)
  $p^0=i\omega_n$ by $\omega+i\eta$,
with real $\omega$ and $\eta\to 0_+$. 
In order to perform the Matsubara sum in eq.~(\ref{Sigeff}),
it is convenient to use the spectral representations
of the various propagators. For $S_0$, this is
given in eq.~(\ref{Sspec0}), with $\rho_{\rm f}(p^0, {\bf p})
= {\slashchar p} \rho_0(p^0, p)$. For the electric and magnetic
 photon propagators we have similarly
\beq\label{Sspec}
{}^*\Delta_t(\omega, {\bf q})&=&\int_{-\infty}^{\infty}\frac{{\rm d}q_0}{2\pi}
\,\frac{{}^*\rho_t(q_0, q)}{q_0-\omega}\,,\nonumber\\
{}^*\Delta_l(\omega, {\bf q})&=&-\frac{1}{q^2}
\,+\int_{-\infty}^{\infty}\frac{{\rm d}q_0}{2\pi}
\,\frac{{}^*\rho_l(q_0, q)}{q_0-\omega}\,,\eeq
where ${}^*\rho_{l}$ and ${}^*\rho_{t}$ are the 
corresponding spectral densities,
\beq\label{rhos}
{}^*\rho_{l,t}(q_0,q) = 2{\rm Im}\,{}^*\Delta_{l,t}(q_0+i\eta ,q)\,.\eeq
Note the subtraction performed in the spectral representation
of ${}^*\Delta_l(\omega, q)$: this is necessary since
${}^*\Delta_l(\omega,q)\to -1/q^2$ as $|\omega|\to \infty$.
When the above expressions are inserted in eq.~(\ref{Sigeff}), the sum
over $\omega_m$ can be performed easily. One obtains then
\beq\label{Sig2}
\Sigma(p)= -\,g^2 \int \frac{{\rm d}^3q}{(2\pi)^3}
\int_{-\infty}^{+\infty}\frac{{\rm d}k^0}{2\pi}
\int_{-\infty}^{+\infty}\frac{{\rm d}q^0}{2\pi}
\,\rho_0(k)\gamma_\mu {\slashchar k} \gamma_\nu\,{}^*\rho^{\mu\nu}(q)\,
\frac{1+N(q^0)-n(k^0)}{k^0+q^0-p^0}\,.\eeq
The analytical continuation   $p^0 \to \omega+i\eta$ can now be done,
and the damping rate is calculated as
   $\gamma(p)= - {\rm Im}\, \Sigma_+(p,p)$.  One gets:
\beq\label{g1}
\gamma(p)&=&\frac{ \pi
g^2}{\omega} \int \frac{{\rm d}^3q}{(2\pi)^3}
\int_{-\infty}^{+\infty}\frac{{\rm d}k^0}{2\pi}
\int_{-\infty}^{+\infty}\frac{{\rm d}q^0}{2\pi}
\delta(k^0+q^0-\omega)\,
\Bigl[1+N(q^0)-n(k^0)\Bigr]\nonumber\\
&\mbox{}&\rho_0(k)\left\{2\Bigl[\omega k_0 - ({\bf p}\cdot \hat{\bf q})
({\bf k}\cdot \hat{\bf q})\Bigr]\,{}^*\rho_t(q)\,+\,
\Bigl[\omega k_0 + ({\bf p}\cdot {\bf k})\Bigr] \,{}^*\rho_l(q)\right\},\eeq
where $\omega=p$, $k^\mu=(k^0, {\bf k})$ and ${\bf k=p-q}$.

 The spectral functions (\ref{rhos}) of the  dressed photon
have the following structure:
\beq \label{RRHO}
{}^*\rho_s(q_0,q)
= 2\pi\epsilon(q_0)\,z_s(q)\,\delta(q_0^2 -\omega_s^2(q))
 +\beta_s (q_0,q) \theta (q^2- q_0^2),\eeq
where $s=l$ or $t$,  $ z_s(q)$ is the residue of the time-like
pole at  ${\omega_s(q)}$, and
\beq\label{RHOLT}
\beta_{l}(q_0,q)&=&3\pi\omega_p^2\,\frac{q_0}{q}\,
|{}^*\Delta_l(q_0,q)|^2,\nonumber\\
\beta_{t}(q_0,q)&=&3\pi\omega_p^2\,\frac{q_0(q^2-q_0^2)}{2q^3}\,
|{}^*\Delta_t(q_0,q)|^2.\eeq
For  $\omega\to p$, the energy conservation 
selects the positive value $k_0= \epsilon_{p-q} \equiv |{\bf p-q}|$
from the  spectral density  $\rho_0(k_0,k)$ of the internal fermion.
Also, the kinematics restricts the photon momentum to be space-like ($|q_0|<q$).
 Finally, because of the infrared sensitivity
of the damping rate, the whole 
contribution to $\Gamma$ in the on-shell limit
  (and not only its divergent part) comes from
{\it soft} photon momenta, $q\ll T$. Since, on the other hand,
  $p \sim T$,
we can make the following kinematical approximations
when evaluating eq.~(\ref{g1}) (recall that $\omega = p\,$):
\beq\label{kinapp}
\epsilon_{p-q} \,\simeq\, p-{\bf p}\cdot \hat {\bf q} &=& p-q\cos \theta,
\nonumber\\
\omega \epsilon_{p-q} - ({\bf p}\cdot \hat{\bf q})
({\bf k}\cdot \hat{\bf q})&\simeq&
p^2(1-\cos^2\theta),\nonumber\\
\omega k_0 + ({\bf p}\cdot {\bf k})&\simeq& 2p^2,\nonumber \\
1+N(q^0)-n(\epsilon_{p-q})&\simeq& N(q^0)\,\simeq\, T/q^0\,.\eeq
With these simplifications,  eq.~(\ref{g1}) becomes
\beq\label{g4}
\gamma(p)&\simeq&
\pi g^2T \int \frac{{\rm d}^3q}{(2\pi)^3}
 \int_{-\infty}^\infty \frac{{\rm d}q_0}{2\pi q_0}
\,\, \delta(q_0 -q\cos \theta)\nonumber\\
&\,&\qquad\qquad \qquad\left({}^*\rho_l(q)+(1-\cos^2\theta)
{}^*\rho_t(q)\right),\eeq
and it is independent of the external momentum.
To be consistent with the approximations performed,
we supply the above integral over $q$ with
an upper cut-off $q^*$ satisfying $gT\ll q^* \ll T$. We shall verify later
that, to the order of interest, the value of the integral
is actually independent of $q^*$.

 By  using  the   $\delta$-function 
 to perform the angular integration in eq.~(\ref{g4}), we obtain
\beq\label{g5}
\gamma &\simeq&
 \frac{g^2T}{4\pi}
\int_{\mu}^{q^*}{\rm d}q \,q\,
 \int_{-q}^{q}\frac{{\rm d}q_0}{ 2\pi q_0}\, 
\left\{\beta_l(q_0,q)+\left(1-\frac{q_0^2}{q^2}\right)
\beta_t(q_0,q)\right\}.\eeq
In order to regularize the IR divergence, we have inserted
a lower cut-off $\mu$ in the integral over $q$.
Note that, because of the kinematics,
 the support of the energy integral is limited to $-q < q_0 < q$, 
so that  only the off-shell pieces
 $\beta_{l,\,t}(q_0,q)$ of the photon spectral densities
 contribute to the damping rate. This is consistent
with the physical interpretation of the damping rate
 presented in section 2.1.  In fact, at this point,
we can easily make contact between these two presentations.
Namely, eq.~(\ref{G2L}) in section 2.1 is essentially the same as 
the above eq.~(\ref{g5}), as can be seen
by using eq.~(\ref{RHOLT}) for the spectral densities.
Moreover, the IR singular piece of the 
damping rate (\ref{g5}) is given
 by eq.~(\ref{G2LR}), as we verify now through a
different computation,  based on the sum-rules \cite{Pisarski93} 
 displayed in Appendix A.

Using the behaviour of these sum-rules for large photon momenta $q\gg \omega_p$,
as given in eq.~(\ref{SumRUV}),  one can verify that
$\gamma$ is independent of the arbitrary
intermediate scale $q^*$, to the order of interest (the contribution
of the momenta $q>q^*$ is of relative order $gT/q^*$).
Furthermore,  the infrared behavior is dominated by
that term of  eq.~(\ref{g5}) which involves
the transverse spectral density divided by $q_0$. Specifically,
 for small momenta $q\ll \omega_p$ we can write
\beq\label{sumt}
\int_{-q}^q\frac{{\rm d}q_0}{2\pi q_0}\,
\beta_t(q_0,q)& = &\frac{1}{q^2}\, \Bigl(1 + {\cal O}(q^2/\omega_p^2)\Bigr),\eeq
which  diverges as $1/q^2$ in the zero momentum limit.
All the other terms give finite contributions as $q\to 0$
(of relative order $q^2/\omega_p^2$), and will be neglected here.
By retaining  only the leading term  in eq.~(\ref{sumt}), we obtain
the singular contribution to eq.~(\ref{g5}):
\beq\label{gammap1}
\gamma_{sing}=
 \frac{g^2T}{4\pi}
\int_{\mu}^{\omega_p}{\rm d}q \,
\,\frac{1}{q}\,=\, \frac{g^2T}{4\pi}\,
\ln \frac{\omega_p}{\mu}.\eeq
The upper cut-off $\omega_p\sim gT$ 
 accounts approximately for the terms which have been
neglected when keeping only the $1/q^2$ contribution to the
 sum-rule (\ref{sumt}) (recall that the full integrand
in eq.~(\ref{g5}) is indeed cut-off at $q\sim \omega_p$).
As long as we are interested only in
the coefficient of the logarithm,
 the precise value of this cut-off is unimportant. The scale $\omega_p$ however
is uniquely determined by the physical process responsible for the existence
of space like photons, i.e., the Landau damping. As we
shall see later, this is the scale which fixes
the long time behavior of the retarded propagator.
\begin{figure}
\protect \epsfysize=20.cm{\centerline{\epsfbox{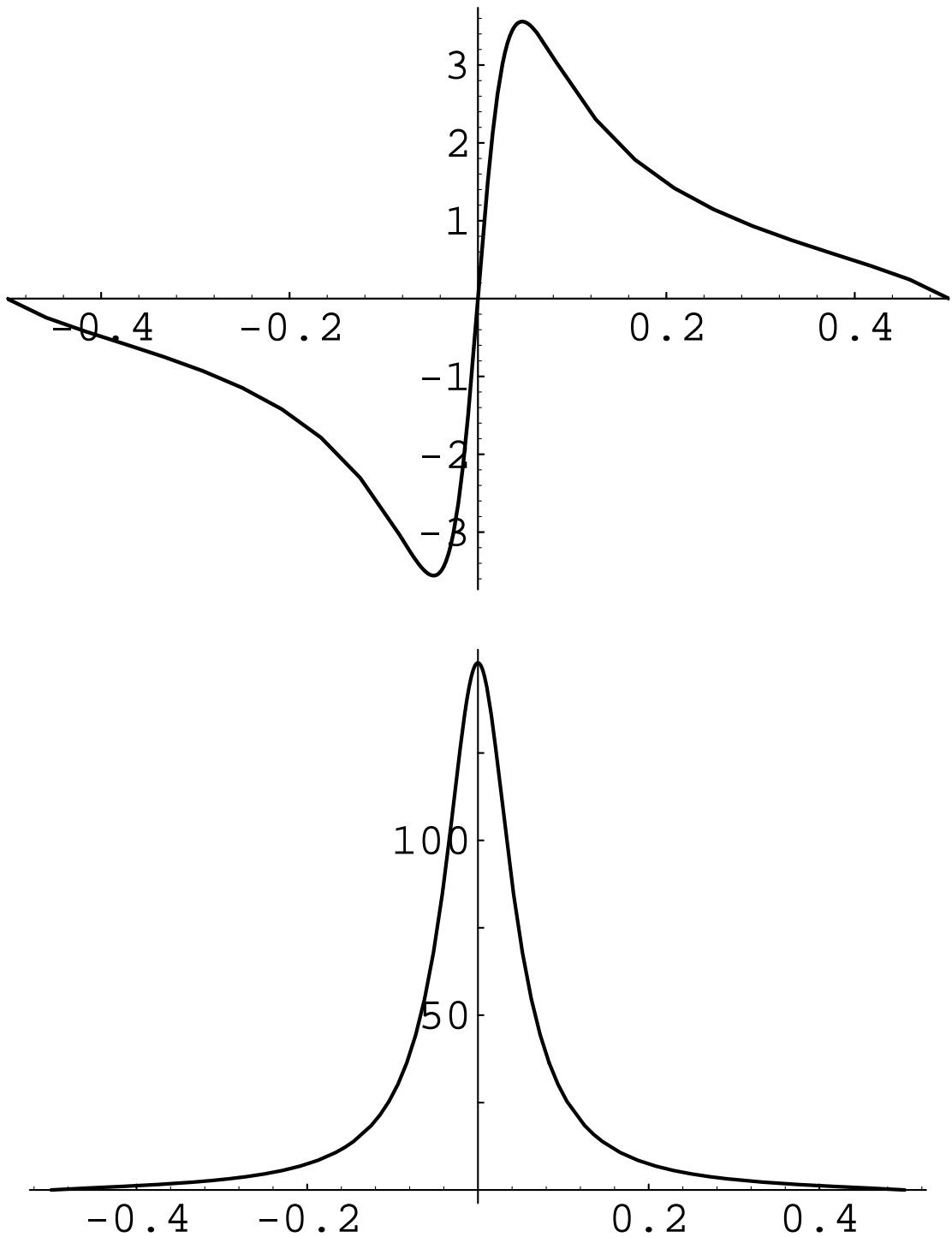}}}
	 \caption{The functions $\beta_t(q_0,q)$ and $\beta_t(q_0,q)/q_0$ 
 for $q=0.5\, \omega_p$.  All the quantities are made adimensional
by multiplying them by appropriate powers of $\omega_p$.}
\label{rho2}
\end{figure}

In terms of collisions, the logarithmic singularity of $\gamma$,
eq.~(\ref{gammap1}), arises from the exchange of very soft
quasistatic ($q_0\simeq 0$) magnetic photons,
as already discussed in section 2.1. 
In the present computation, this may be seen also  as follows:
for very soft momenta $q\ll \omega_p$, the function $\beta_t(q_0,q)/q_0$ 
is strongly peaked at $q_0=0$ (see Fig. \ref{rho2})
 and in the calculation
of the integral (\ref{sumt}) it can be replaced by the following
approximate expression:
\beq\label{rhot}
\frac{1}{q_0}\,\beta_t(q_0\ll q)= \frac{3\pi}{2}\, \frac{\omega_p^2 \, q}
{q^6\,+\, (3\pi \omega_p^2 q_0/4)^2}\,\longrightarrow_{q\to 0}
\,\frac{2\pi}{q^2}\,\delta(q_0).\eeq
This is, of course, just a translation of the corresponding
property (\ref{singDT}) of the magnetic propagator.
Still, this is suggestive as it shows that the
 density of states $\beta_t(q_0,q) N(q_0) \sim (T/q_0) 
\beta_t(q_0,q)$ which are available for the emission ($q_0>0$)
or the absorbtion ($q_0<0$) of a virtual photon with momentum
${\bf q}$ and energy $q_0$ is {\it nonvanishing}
in the zero-frequency limit $q_0\to 0$, in spite of the fact that
the spectral density $\beta_t(q_0,q)$ vanishes in the same limit.
In fact, for very soft momenta $q\ll \omega_p$,
the whole density of states is concentrated at $q_0=0$,
as shown by eq.~(\ref{rhot}).

\subsection{Static photon modes and non-perturbative aspects}

Because of the delta function singularity $\delta(q_0)$ in
eq.~(\ref{rhot}), the above discussion suggests that,
in the imaginary time formalism,  the whole IR singularity is
concentrated in the static mode  $q_0=0$.
Let us verify this explicitly by showing that, indeed,
the logarithm in eq.~(\ref{gammap1})
arises entirely from the magnetic contribution of
the {\it static} term $q_0=i\omega_m=0$  in the Matsubara sum 
of eq.~(\ref{Sigeff}) \cite{Burgess92,Rebhan95}.
 Note that the analytic continuation of this term
to real energy ($p_0\to \omega+i\eta$) is well-defined,
since all its singularities lie on the real axis
 in the complex $p_0$ plane.
(This is not so for the terms with  $q_0=i\omega_m\ne 0$,
which individually have singularities off the real axis.)

The   magnetostatic mode gives the
following contribution to the one-loop self-energy:
\beq\label{Sig0}
\Sigma_s(\omega,{\bf p})&=&-\,g^2 T
 \int \frac{{\rm d}^3q}{(2\pi)^3}\,
\,\gamma^i\,S_0(\omega, {\bf p-q})\,\gamma^j
\,{}^*D^{ij}(0,{\bf q})\,\nonumber\\
&=&\,g^2 T \int \frac{{\rm d}^3q}{(2\pi)^3}\,\frac
{\gamma^i(\omega\gamma^0- ({\bf p-q})\cdot \bfgamma)\gamma^j}
{(\omega+i\eta)^2-({\bf p-q})^2}\,\frac{\delta^{ij}- 
\hat q^i \hat q^j} {{\bf q}^2}\,.\eeq
The momentum integral in eq.~(\ref{Sig0})
shows a logarithmic {\it ultraviolet} divergence.
In the full calculation, such a divergence would be cut-off
by the contribution of the non static modes. 
(Recall the discussion after eq.~(\ref{gammap1}).)
When supplemented with an upper cut-off $\omega_p$,
eq.~(\ref{Sig0}) yields the following
contribution to the fermion damping rate (for $\omega\simeq p$):
\beq\label{gammaIR}
\gamma_s
 &\equiv& - \frac{1}{4p}\,{\rm tr}\left(
{\slashchar p}\,{\rm Im}\, \Sigma_s(p)\right)
\simeq  g^2  T\int \frac{{\rm d}^3q}{(2\pi)^3}\,\frac{1}{q^2}\,
\, {\rm Im}\,\,\frac
{-1}{\omega -p -q\cos\theta +i\eta}\nonumber\\& \simeq &
 \frac{g^2T}{4\pi}
\int_{0}^{\omega_p} {\rm d}q \int_{-1}^1 {\rm d}\cos\theta\,
\delta(\omega -p -q\cos\theta)\,=\,\alpha T\ln \frac{\omega_p}{|\omega-p|},
\eeq
where the approximate equality means that only regular terms have 
been dropped.  In the mass-shell limit $\omega\to p$,
 this reproduces the singular result of eq.~(\ref{gammap1}).
Note that the upper cut-off $\omega_p$ is the only trace of the screening
effects in the above calculation: indeed, the magnetostatic propagator
is the same as at the tree-level, namely ${}^*\Delta(0, q)=1/q^2$.

The divergence of $\gamma$ at the (resummed) one-loop level
invites a closer examination of the higher order corrections. 
The two-loop self-energy is briefly discussed in Appendix C,
where we show that the leading infrared divergence arises,
again from terms where both the internal photons are static and magnetic.
This result is readily generalized to all orders:
the most singular contributions to the
on-shell fermion self-energy are confined to the magnetostatic sector.
When computing these contributions, all the loop integrals run over the
three momenta ${\bf q}$ of the static internal photons, so that the
infrared  singularities are effectively those of a
 {\it three-dimensional} theory. 
Consider then a generic  $n$-loop self-energy diagram with
 only magnetostatic modes:
Its discontinuity, when evaluated on the tree-level mass-shell $\omega=p$,
has power-like IR divergences, possibly combined with logarithmic ones.
Power counting shows that the leading divergences
are of relative order $(g^2T/\mu)^{n-1}$, where $\mu$ is an IR cut-off.
Such  strong IR divergences are analogous to those identified in the
analysis of the corrections to the screening mass in \cite{debye}, and their
presence signals a breakdown of perturbation theory. 

To get further insight, it is useful to consider
the explicit two-loop calculation from Appendix C: 
the on-shell self-energy $\Sigma^{(2)}(p, {\bf p})$
shows a linear plus logarithmic divergence.
(There are also subleading, purely logarithmic,
divergences, but these are left out in a leading-order
calculation.) Specifically,  eq.~(\ref{PIS1}) yields
\beq\label{Sigma2L}
\Sigma_+^{(2)}(p,p) \equiv \frac{1}{2}\,{\rm tr}\Bigl(
h_\pm(\hat {\bf p}) \Sigma^{(2)}(p,p)\Bigr)
\,\simeq\,i \,\frac{2}{\pi}\,\frac{(\alpha T)^2}{\mu}\,\ln
\frac{\omega_p}{\mu}\sim \frac{\alpha T}{\mu}\,
\Sigma_+^{(1)}(p,p)\,,\eeq
where $\Sigma_+^{(1)}(p,p)= - i \alpha T\ln(\omega_p/\mu)$
is the on-shell limit of the one-loop self-energy in eq.~(\ref{gammaIR}).
Now, in order to compute the two-loop contribution to the damping rate,
one has to expand the dispersion equation $\omega = p+ \Sigma_+(\omega, p)$
up to the order of interest. This yields the 2-loop mass-shell correction 
in the form 
\beq\label{domeg}
 \delta\omega^{(2)}(p) = [z^{(1)}(p)-1] \Sigma_+^{(1)}(p,p) +
 \Sigma_+^{(2)}(p,p) + {\cal O}(3 \,\,{\rm loops})\,,\eeq
where
\beq\label{z1}
 z^{(1)}(p) - 1 = \frac{\del \Sigma_+^{(1)}}{\del\omega}\bigg |_{\omega=p}
\simeq \,\frac{2}{\pi}\,\frac{\alpha T}{\mu} \eeq
is the one-loop residue, whose leading IR-divergent part has been
 computed in the Appendix. By combining eqs.~(\ref{Sigma2L}) and
(\ref{domeg})--(\ref{z1}), we note that the leading, power-like,
divergences cancel between the two-loop self-energy and the one-loop
residue, so that the two-loop correction to $\gamma$ is only
logarithmically divergent, as at the one loop level. 

A simple argument, based on a gauge-invariant approximation to
the full Dyson-Schwinger equation which is detailed in Appendix C,
 suggests that this is a general feature: 
{\it if we assume the fermion propagator to have
a simple pole at the mass-shell}, then
the damping rate remains {\it logarithmically} divergent 
 to {\it all} orders. That is, the 
power-like divergences which occur in $\Sigma (\omega=p)$ 
appear to cancel
against similar divergences in the residue. (A similar all-order
cancellation has been argued in three-dimensional
QED at zero temperature\cite{Sen}.) However,
the persistence of the logarithmic divergence in all
orders of perturbation theory suggests that the analytic
structure of the propagator is more complicated than a simple pole.

To conclude this section, let us emphasize that
when we compute the imaginary part of multi-loop diagrams with
only static internal photons, we are actually considering
 the effects of multiple collisions involving the
 exchange of quasistatic magnetic photons with the plasma particles.
The fact that these processes (or, more accurately,
their most IR singular contributions to the interaction rate)
 can be effectively taken into
account by the ``dimensional reduction'' to the
magnetostatic photon modes  is  a
consequence of the specific infrared behaviour of the resummed
magnetic propagator, as expressed by eq.~(\ref{singDT}).

\section{The Bloch-Nordsieck model at finite temperature}
\setcounter{equation}{0}

Previously, we have shown that the leading infrared divergences in
the perturbative computation of the fermion self-energy
are those of an effective three-dimensional theory
involving only static magnetic photons.
 We shall take advantage of this in order to get
an explicit expression for the fermion propagator.
 However, before restricting ourselves
to the static photon modes, we shall first
develop a more general approach which
 is essentially a finite-temperature extension
of the Bloch-Nordsieck approximation\cite{BN37,Bogoliubov}.

\subsection{Perturbation theory with soft photons}

We start by deriving a set of simplified Feynman rules 
which allows one to compute the most IR singular contributions
to the damping rate from higher loop self-energy diagrams.
The leading  infrared divergences arise from diagrams where
{\it all} the internal photon lines are soft, 
and therefore dressed by the screening corrections.
No further resummation of the photon lines is necessary
beyond the hard termal loop approximation: in abelian gauge theories,
 all the higher order corrections to the photon polarisation
tensor remain perturbative, and do not modify the qualitative
 IR behavior of the HTL-resummed propagator (denoted
as ${}^*D_{\mu\nu}(q)$).

\begin{figure}
\protect \epsfxsize=16.cm{\centerline{\epsfbox{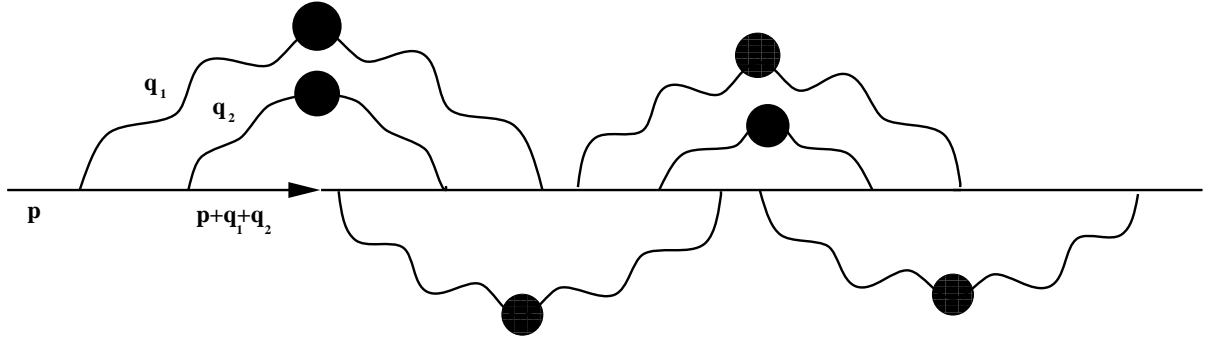}}}
	 \caption{A generic $n$-loop diagram (here, $n=6$)
for the self-energy in quenched QED.}
\label{Nloop}
\end{figure}

Thus, when expressed in terms of the resummed photon propagator,
 the relevant self-energy diagrams contain no fermion loops:
 the internal photon lines are all attached on
the incoming fermion line. A typical $n$-loop diagram
is shown in Fig.~\ref{Nloop}. There are as many loops
as photon propagators, and we can chose all the
independent loop momenta to be the momenta
$q_r$ of the soft photon lines
(here $r=1,\,...,n$ for an $n$-loop graph). 
All such diagrams are composed from the three following
structural units:\\
(i) the effective photon propagator  ${}^*D_{\mu\nu}(q)$;\\
(ii) the fermion propagator $S_0(p+q)$, where $p$ is the hard external
momentum, and $q$ is a linear combination of the soft loop momenta;\\
(iii) the photon-fermion vertex $\gamma^\mu$.\\
In the kinematical regime of interest, both
the  fermion propagator and the vertex function can
be further simplified, along the lines explained in Section 2.2.
After performing the Matsubara sums over the internal
bosonic frequencies, and the analytic continuation
to real external energy, the internal fermion lines
are represented by spectral densities like (see, e.g., eq.~(\ref{Sig2}))
\beq \label{r1}
\rho_{\rm f}(k^0, {\bf p-q})\,=\,\Bigl(k^0\gamma^0 -({\bf p-q})\cdot {\bfgamma}
\Bigr)\frac{\pi}{\epsilon_{p-q}}\,\Bigl(\delta(k^0-\epsilon_{p-q})-
\delta(k^0+\epsilon_{p-q})\Bigr), \eeq 
which multiply energy denominators of the form $1/(\omega-k^0-q^0)$. 
Since $q\ll p$, we can use $\epsilon_{p-q}\simeq \epsilon_p-
{\bf v}\cdot {\bf q}=p-q\cos \theta$ (where
${\bf v}=\del \epsilon_p/\del {\bf p} = \hat{\bf p}$)
to replace eq.~(\ref{r1}) with
\beq\label{r2}
\acute\rho_{\rm f}(k^0, {\bf p-q})\equiv 
(\gamma^0 - \hat{\bf p}\cdot {\bfgamma})
{\pi}\delta(k^0- {\bf v}\cdot {\bf (p-q)}) = h_+(\hat{\bf p})
\acute\rho_0(k^0, {\bf p-q}),\eeq
where the reduced spectral density
\beq\label{newrho}
\acute\rho_0(\omega, {\bf p})\equiv 2\pi\delta(\omega
-{\bf v}\cdot {\bf p}) \eeq
involves only the positive-energy fermion state.
The contribution of the negative-energy fermion
state, initially present in eq.~(\ref{r1}), is
 suppressed by the corresponding large energy denominator.

One sees on eq.~(\ref{r2}) that neither
the spin structure, nor the negative-energy fermion  intermediate
 states, play an important role. In fact, the
residual spin structure of eq.~(\ref{r2}), i.e.
the spin matrix $h_+(\hat{\bf p})$, does not involve the
loop momenta anymore, and can be absorbed into a redefinition
of the vertex function. To see this, recall that, for
a positive energy hard fermion, the relevant self-energy
is $\Sigma_+= {\rm tr}\,(h_+(\hat {\bf p})\Sigma)/2$. 
In the present kinematical regime, the spin
structure of a typical $n$-loop contribution to $\Sigma_+$ factorizes
  into the trace
\beq 
I^{\mu_1\mu_2\,...\,\mu_{2n}}\,=\,\frac{1}{2}\,
{\rm tr}\,\Bigl\{h_+(\hat {\bf p})\gamma^{\mu_1}
h_+(\hat {\bf p})\gamma^{\mu_2}\,...\,
h_+(\hat {\bf p})\gamma^{\mu_{2n}}\Bigr\}.\eeq
By using the identities (with $v^\mu=(1, {\bf v})$)
\beq\label{IDh}
h_+(\hat {\bf p})\gamma^{\mu}h_+(\hat {\bf p})&=& v^\mu
h_+(\hat {\bf p}),\nonumber\\
{\rm tr}\,\Bigl(h_+(\hat {\bf p})\gamma^{\mu}\Bigr)&=& 2v^\mu,\eeq
one readily derives
\beq 
I^{\mu_1\mu_2\,...\,\mu_{2n}}\,=\,v^{\mu_1}\,v^{\mu_2}\,...\,
v^{\mu_{2n}}\,.\eeq
The same result would have been obtained
by using the reduced spectral density (\ref{newrho})
instead of (\ref{r2}), together with the effective
vertex $\Gamma^\mu=v^\mu$.

To conclude, the  corrections to the self-energy $\Sigma_+$ which derive from 
the fermion interactions with soft photons can be obtained
from the Feynman graphs of quenched QED, by evaluating the
latter with the following effective Feynman rules:\\
(i) the photon propagator   ${}^*D_{\mu\nu}(q)$;\\
(ii) the fermion (analytic) propagator 
\beq\label{G0}
G_0(p-q)\,=\,\int_{-\infty}^{+\infty}\frac{{\rm d}k^0}{2\pi}\,
\frac{\acute\rho_0(k^0, {\bf p-q})}{k^0-(p^0-q^0)}
\,=\,\frac{-1}{(p^0-q^0)-{\bf v}\cdot ({\bf p-q})}\,;\eeq
(iii) the photon-fermion vertex $\Gamma^\mu=v^\mu$.\\
Any reference to the spin structure, 
and also to the antiparticles, has disappeared.  

We note that,  when used in relation to the one-loop self-energy
in Fig.~4, the above Feynman rules
 yield directly the expression (\ref{g5}) for the damping rate
(that is, the {\it whole} contribution of order $g^2T$, and not
only its divergent piece).
For higher loop diagrams however, we do not expect all the subleading 
divergences  to be correctly reproduced since,
for instance, contributions coming from mixed diagrams,
where some photons are hard and the other ones are soft,
have been ignored.

The simplified structure which is put forward here
 is familiar from most treatements
of the IR divergences at zero temperature (see, e.g.,
\cite{Weinberg} and references therein). It
 can be most economically exploited 
within  the {\BN} model\cite{BN37} (see also \cite{Bogoliubov}), which,
for the vaccuum theory, is exactly soluble.
In order to search for a non-perturbative solution at finite
temperature, we follow Ref. \cite{Bogoliubov} and
reformulate this model in the language of path integrals. 

\subsection{The Bloch-Nordsieck model in functional form}

In the Matsubara formalism,
the  exact fermion propagator at finite temperature can be obtained as
\beq\label{FunctionalS}
S_E(x,y)= Z^{-1}\int [{\rm d} A]\, S_E(x,y|A)\,{\rm exp}\left\{-
{\rm Tr}\,{\rm ln}\,  S_E(x,y|A)-
\frac{1}{2}\,\Bigl(A,D_0^{-1}A\Bigr)\right\}\,\,\eeq
 where  $ S_E(x,y|A)$ is the (imaginary-time)
fermion propagator in the presence
of a background gauge field:
\beq\label{SA}
-i{\slashchar D}_x  S_E(x,y|A)=\delta_E (x-y),\eeq
and $D_\mu=\del_\mu +igA_\mu$. In these equations, the time
variables are purely imaginary (e.g., $x_0=-i\tau_x$ and $y_0=-i\tau_y$,
with $0\le \tau_x,\, \tau_y \le \beta$ and
$\delta_E(x-y)= \delta(\tau_x-\tau_y)\delta({\bf x-y})$),
and the gauge fields are periodic in time, $A_\mu(\tau=0)=A_\mu(\tau=\beta)$.
The tree-level photon action has been written as
\beq\label{SEFF} 
\Bigl(A,D_0^{-1}A\Bigr)\,=\,T\sum_m \int \frac{{\rm d}^3q}{(2\pi)^3}\,
A^\mu(i\omega_m, {\bf q})\, D^{-1}_{0\,\mu\nu} (i\omega_m, {\bf q})
A^\nu (-i\omega_m, -{\bf q}),\eeq
where $\omega_m=2\pi mT$ with integer $m$,
 and $D_{0\,\mu\nu}$ is the free photon propagator
 in an arbitrary gauge.

The fermion propagators $S_E(x-y)$ and $S_E(x,y|A)$ are antiperiodic.
For instance,
\beq\label{BC} S_E(\tau_x=0,\tau_y|A)=- S_E(\tau_x=\beta,\tau_y|A),\eeq
and similarly for $\tau_y$.
 The functional determinant ${\rm exp}\{-{\rm Tr}\,{\rm ln}\,S_E(x,y|A)\}$ 
describes the plasma polarization. Diagrammatically, this terms
generates internal fermion loops. As already discussed,
 the only polarization effects which need
to be considered are those contained in the photon HTL,
which we denote here as $\delta\Pi_{\mu\nu}$:
\beq
{\rm Tr}\,{\rm ln}\,  S_E(x,y|A)\simeq
\frac{1}{2}\,\Bigl(A,\, \delta\Pi\,A\Bigr).\eeq
Furthermore, the  simplifications discussed  in the previous subsection
are easily implemented by replacing the exact propagator $S_E(x,y|A)$ 
in eq.~(\ref{FunctionalS}) with the Bloch-Nordsieck propagator
$G_E(x,y|A)$, solution of the equation
\beq\label{GA}
-i\,(v\cdot D_x)\,G_E(x,y|A)&=&\delta_E(x-y),\eeq
with antiperiodic boundary conditions. 
(Formally, this equation is obtained by replacing
the Dirac matrices $\gamma^\mu$ by the particle velocity
$v^\mu$ in the full equation (\ref{SA})).
With the above simplifications, the general  equation (\ref{FunctionalS})
reduces to 
\beq\label{FS0}
S_E(x,y)= Z^{-1}\int [{\rm d} A]\, G_E(x,y|A)\,{\rm exp}\left\{-
\frac{1}{2}\,\Bigl(A,\, {}^*D^{-1}A\Bigr)\right\},\eeq
with ${}^*D^{-1}_{\mu\nu}=D_{0\,\mu\nu}^{-1}+\delta\Pi_{\mu\nu}$.
It is easy to verify that, when considered 
in perturbation theory, eqs.~(\ref{GA}) and (\ref{FS0})
generate  the simplified
Feynman rules alluded to at the end of the previous subsection.

\subsection{The Bloch-Nordsieck propagator in imaginary-time}

In real-time,  the  equation for $ G(x,y|A)$ reads
\beq\label{Gret}
-i\,(v\cdot D_x)\,G(x,y|A)&=&\delta^{(4)}(x-y),\eeq
and can be solved  {\it exactly}. For retarded boundary conditions,
$ G(x,y|A) =0 $ for $x_0<y_0$, the solution reads
\beq\label{GR}
G_{R}(x,y|A)&=&i\,\theta (x^0-y^0)\,\delta^{(3)}
\left({{\bf x}}-{{\bf y}}-{{\bf v}}(x^0-y^0)
\right )U(x,y)\nonumber\\
&=&i\,\int_0^\infty {\rm d}t\,\delta^{(4)}(x-y-vt)\,U(x,x-vt).\eeq
Here,  $U(x,y)$  is the parallel transporter along the straight line
trajectory of velocity ${\bf v}$ joining $x$ and $y$ ($y=x-vt$):
\beq\label{U}
U(x,x-vt)=\exp\left\{ -ig\int_0^t {\rm d}s \, v\cdot A(x-v(t - s))
\right\}.\eeq
In order to verify that (\ref{GR}) is indeed a solution
of eq.~(\ref{Gret}), one may use the fact
that the function $U(x,x-vt)$ satisfies the following equation,
\beq\label{derU}
-\frac{\del}{\del t}\,U(x,x-vt)= (v\cdot D_x)\,U(x,x-vt),
\eeq
with the boundary condition $U=1$ for $t=0$.

In imaginary-time, the resolution of eq.~(\ref{GA}) is complicated
by the antiperiodic boundary conditions to be imposed on $G_E$:
\beq\label{BCG} G_E(\tau_x=0,\tau_y|A)=- G_E(\tau_x=\beta,\tau_y|A),\eeq
and similarly for $\tau_y$. The free equation ($A=0$) can be easily
solved  in momentum space:
\beq\label{GEfree}
G_E(i\omega_n,{\bf p})&=&\frac{1}{{\bf v\cdot p} - i\omega_n},\eeq
where $\omega_n=(2n+1)\pi T$. This is in agreement with
eq.~(\ref{G0}). In the  imaginary time representation,
\beq\label{GOtau}
G_E(\tau,{\bf p})\,=\,\sum_{\omega_n}
{\rm e}^{-i\omega_n\tau}\,G_E(i\omega_n,{\bf p})\,=\,
{\rm e}^{- {\bf v\cdot p}\tau}\,\Bigl[\theta(\tau)(1-n({\bf v\cdot p}))
\,-\,\theta(-\tau)n({\bf v\cdot p})\Bigr],\eeq
where $n(\omega) = 1/\Bigl({\rm exp}(\beta\omega)+1\Bigr)$
is the Fermi-Dirac statistical factor.


Consider now the interacting problem, with $A\ne 0$.
As a guidance in searching a solution to eq.~(\ref{GA}) with
antiperiodic boundary conditions, we use
 the solution  (\ref{GR}) to the real-time problem, 
 which we write in the form (with $p^\mu=(\omega, {\bf p})$
and $v\cdot p= \omega - {bf v\cdot p}$)
\beq\label{GR1}
G_R(x,y|A)&=&\int\frac{{\rm d}^4p}{(2\pi)^4}
\,{\rm e}^{-ip\cdot(x-y)}G_R(x,p|A)
\nonumber\\
G_R(x,p|A)&=&i\int_0^\infty {\rm d}u\,{\rm e}^{it(v\cdot p)-\eta t}
U(x,x-vt),\eeq
where the $x$-dependence of the function
$G_R(x,p|A)$ comes from the corresponding dependence of the background field.


By analogy, we look for the solution
$G_E(x,y|A)$ to the imaginary-time BN equation in the following
 form\footnote{To simplify notations, the measure in the momentum integrals
 will be denoted below by the following condensed notation:
$$\int[{\rm d}q]\equiv T\sum_{q_0, even}
 \int\frac{{\rm d}^3q}{(2\pi)^3}\,,
\qquad
 \int\{{\rm d}p\}\equiv  T\sum_{p_0, odd} \int\frac{{\rm d}^3p}{(2\pi)^3}\,.$$}
\beq\label{ansatz}
G_E(x,y|A)&=&\int \{{\rm d}p\}\,{\rm e}^{-ip\cdot(x-y)}G_E(x,p|A)
\nonumber\\
G_E(x,p|A)&=&-\int_0^\beta {\rm d}u\,{\rm e}^{-u(v\cdot p)}
V(x,{\bf v\cdot p};u),\eeq
where the unknown function $V(x,{\bf v\cdot p};u)$
satisfies 
\beq\label{eqV}
-\frac{\del}{\del u}\, V&=&i(v\cdot D_x)V,\nonumber\\
V(\tau_x=0;u)&=&V(\tau_x=\beta;u),\nonumber\\
V(x,{\bf v\cdot p};u=0) &+&
{\rm e}^{\beta ({\bf v\cdot p})}V(x,{\bf v\cdot p};u=\beta)=1.\eeq
As in the real-time case, the $x$-dependence of $G_E(x,p|A)$ arises entirely
from its interactions with the (periodic) gauge field. If $A=0$,
we recover the free propagator (\ref{GEfree}) by replacing
 $V(x,{\vp};u)$ with $n({\vp})$,
which  satisfies indeed the last equation (\ref{eqV})
because of the identity
\beq\label{ID1}
n(\epsilon)+ {\rm e}^{\beta\epsilon}n(\epsilon)=1.\eeq

Eq.~(\ref{eqV}), with the indicated boundary conditions,
can be solved as a series in powers of $gA_\mu$, that is,
 as a perturbative expansion:
\beq\label{Vpert}
V(x,{\vp};u)&=&n({\vp})+g\int[{\rm d}q]\,{\rm e}^{-iq\cdot x}\,
\frac {v\cdot A(q)}{v\cdot q}\,\Bigl[n({\vp})- n({\vpq}){\rm e}^{-u(v\cdot q)}
\Bigr]\nonumber\\
&+&\frac{g^2}{2}\int[{\rm d}q_1][{\rm d}q_2]\,{\rm e}^{-i(q_1+q_2)\cdot x}
\frac {v\cdot A(q_1)}{v\cdot q_1}\,
\frac{{\tilA}(q_2)}{v\cdot q_2}\,\nonumber\\
&{}&\Bigl[n({\vp})-n({\bf v\cdot (p+ q_1)})
\,{\rm e}^{-u(v\cdot q_1)}-n({\bf v\cdot(p+ q_2)})\,{\rm e}^{-u(v\cdot q_2)}
\nonumber\\
&{}&+n({\bf v\cdot(p+ q_1+q_2)} )\,{\rm e}^{-u\,v\cdot( q_1+ q_2)}\Bigr]
+\, ... .\eeq
It can be verified, using in particular
the identity (\ref{ID1}), that
the series (\ref{Vpert}) satisfies indeed eqs.~(\ref{eqV}).

As already noted, the quantity $V(x,{\vp};u)$ is the imaginary-time analogue of
the real-time parallel transporter $U(x,x-vt)$, eq.~(\ref{U}).
This is also manifest from the analogy between
eq.~(\ref{derU}) for $U(x,x-vt)$ and  eq.~(\ref{eqV})
for  $V(x,{\vp};u)$. By solving eq.~(\ref{derU})
 in perturbation theory, one
generates a series analogous to (\ref{Vpert}),
where, however, the thermal factors are absent.
The correspondance between the two series can be easily worked out
term by term.  For instance,
$$\Bigl[n({\vp})-n({\vpq}){\rm e}^{-u(v\cdot q)}
\Bigr]\to \Bigl(1-{\rm e}^{it(v\cdot q)}\Bigr)\,,$$
and so on. In the real-time series, factorisations occur,
which bring in simplifications. For example, in second order,
$$ \Bigl(1-{\rm e}^{it(v\cdot q_1)}-{\rm e}^{it(v\cdot q_2)}
+{\rm e}^{it(v\cdot q_1+v\cdot q_2)}\Bigr)=
 \Bigl(1-{\rm e}^{it(
v\cdot q_1)}\Bigr)\Bigl(1-{\rm e}^{it (v\cdot q_2)}\Bigr)\,.$$
Because of such factorisations,
the real-time series corresponding to eq.~(\ref{Vpert}) can
be resummed into an exponential, leading to the expression (\ref{U}).
In the imaginary-time, the presence of the thermal
factors prevents such a simple exponentiation.

By inserting $G_E(x,y|A)$, eqs.~(\ref{ansatz}) and (\ref{Vpert}),
into eq.~(\ref{FS0}), we can perform
the gaussian functional integral over the photon
fields term by term. This yields:
\beq\label{SE}
S_E(x-y)&=&\int \{{\rm d}p\}\,{\rm e}^{-ip\cdot(x-y)}S_E(p)
\nonumber\\
S_E(p)&=&-\int_0^\beta {\rm d}u\,{\rm e}^{-u(v\cdot p)}
\,\tilde V({\bf v\cdot p};u),\eeq
where $\tilde V({\vp};u)$ is the functional average of
$V(x,{\vp};u)$, eq.~(\ref{Vpert}), 
\beq\label{tilVpert}
\tilde V({\vp};u)&=&n({\vp})+
\sum_{n\ge 1}(-1)^n\frac{g^{2n}}{n!}
\int[{\rm d}q_1 {\rm d}q_2\,...\,{\rm d}q_n]
\frac{\tilde D(q_1) \tilde D(q_2)\,...\,\tilde D(q_n)}
{(v\cdot q_1)^2 (v\cdot q_2)^2\,...\,(v\cdot q_n)^2}\nonumber\\
&{}&\Bigl[n({\vp})-n({\bf v\cdot (p +q_1)})\,
{\rm e}^{-u(v\cdot q_1)}-n({\bf v\cdot (p+q_2)})\,{\rm e}^{-u(v\cdot q_2)}
+\nonumber\\ &{}&\,...\,+(-1)^n\, n({\bf v\cdot (p +q_1+q_2+...+q_n)})\,
{\rm e}^{-u\,v\cdot(q_1+ q_2+...+ q_n)}\Bigr]
\,,\eeq
and
\beq\label{tilD}
\tilde D(q)&=&v^\mu\, {}^*D_{\mu\nu}(i\omega_m,{\bf q}) v^\nu\,.\eeq
Eqs.~(\ref{SE})--(\ref{tilVpert}) express
 the Matsubara fermion propagator in the {\BN} model
as a formal series in powers of $g^2$.

\subsection{The retarded propagator}

To study the mass-shell behavior of the fermion propagator,
we need the {\it retarded} propagator, rather than the Matsubara one.
These two propagators are related by analytic continuation
in either the complex energy, or the complex time, plane.
Here it is more convenient to proceed
in the time representation. To this aim, we recall that
the retarded propagator $S_R(t,{\bf p})$ can be obtained as
\beq\label{SR}
S_R(t,{\bf p})&=&i\theta(t)\Bigl(S^>(t, {\bf p})+S^<(t, {\bf p})\Bigr),
\eeq
where the functions $S^>$ and $S^<$ are the analytic components
of the time-ordered propagator\cite{KB62,MLB96}.
These can be obtained from the
Matsubara propagator:
\beq\label{SE0}
S_E(\tau,{\bf p})&=&\theta(\tau)S^>(\tau, {\bf p})-
\theta(-\tau)S^<(\tau, {\bf p}).\eeq

In order to get the Matsubara propagator
 we have to evaluate first the sum over
$p_0=i\omega_n$ in eq.~(\ref{SE}). Since $\tilde V({\bf v\cdot p};u)$ 
is independent of $p_0$, this may be done trivially,
by using 
\beq
T\sum_{n, \,odd} {\rm e}^{-i\omega_n(\tau +u)}=
\delta(\tau+u)-\delta(\tau +u-\beta).\eeq
Then, for $-\beta \le \tau\le 0$, we obtain
\beq \label{S<}
S^<(\tau, {\bf x})&=&\int \frac{{\rm d}^3p}{(2\pi)^3}\,
{\rm e}^{i{\bf p}\cdot {\bf x}}\,S^<(\tau, {\bf p}),\nonumber\\
S^<(\tau, {\bf p})&=& {\rm e}^{-\tau({\vp})} \tilde V({\vp}; u=-\tau),\eeq
and similarly,  for $0\le \tau\le \beta$,
\beq\label{S>}
S^>(\tau, {\bf p})&=&
 {\rm e}^{(\beta-\tau)({\vp})} \tilde V({\vp}; u=\beta-\tau).\eeq
In particular, the last eq.~(\ref{eqV}) implies
\beq\label{sum}
S^>(0, {\bf p})\,+\,S^<(0, {\bf p})\,=\,1.\eeq
If the functions $S^<(\tau)$ and $S^>(\tau)$ are known
explicitly, then they can be analytically extended
in the complex time plane by simply replacing $\tau\to it$, with complex $t$.
The functions $S^<(t)$ and $S^>(t)$ thus obtained
are well defined for any $t$ satisfying $0\le {\rm Im}\, t\le\beta$,
in the case of $S^<(t)$, and  $-\beta\le {\rm Im}\, t \le 0$,
for $S^>(t)$. For the problem at hand,
these analytic properties can be verified
on eq.~(\ref{tilVpert}): they arise from  the fact
that the thermal factors render the momentum integrals like
$$\int\frac{{\rm d}^3q}{(2\pi)^3}\,n({\vpq})\,{\rm e}^{u({\vq})}$$
 convergent for any $0< u <\beta$.
We see that the statistical factors are essential to ensure the correct
analytical properties; but, at the same time, they prevent
the exponentiation in  eq.~(\ref{tilVpert}). 

According to eqs.~(\ref{SR}), (\ref{S<}) and (\ref{S>}),
the  retarded propagator is given by
\beq\label{SRE}
S_R(t,{\bf p})=i\theta(t) {\rm e}^{-it({\vp})}\Bigl(
  {\rm e}^{\beta({\vp})} \tilde V({\vp}; u=\beta-it)
+ \tilde V({\vp}; u=-it) \Bigr).\,\eeq
The analytic continuation 
of the function $\tilde V({\vp}; u)$ to real time is permitted only
{\it after} performing the Matsubara sums over the bosonic
frequencies $q^0=i\omega_m$ in all the terms
of the infinite series (\ref{tilVpert}).
Fortunately, we may avoid doing this if 
we restrict ourselves to resumming the
{\it leading} infrared divergences. This is further explained
in the next subsection.

\subsection{Dimensional reduction}

In view of the discussion in section 2.3,
the most IR singular terms of the perturbative expansion
are concentrated in the static photon modes.
 Considering only the contribution
of the static modes  $q_0=i\omega_m=0$ to
eq.~(\ref{tilVpert}) is equivalent to solving
the Bloch-Nordsieck equation (\ref{GA}) in the presence
of a {\it static} background field $A_\mu({\bf x})$:
\beq\label{Ast} A^\mu({\bf x}) = T\int \frac{{\rm d}^3q}{(2\pi)^3}\,
{\rm e}^{i{\bf q\cdot x}}A_{\mu}(\omega_m=0,{\bf q})
= T \int_0^\beta {\rm d}\tau\,A_{\mu}(\tau,{\bf x}).\eeq
With only static photon modes, the analytic continuation
of eq.~(\ref{tilVpert}) to real time is trivial,
and the sum in eq.~(\ref{SRE}) can be performed explicitly,
term by term. As we show now, the thermal occupation factors
{\it compensate in this sum}, and the resulting series
 for $S_R(t,{\bf p})$  can be
resummed as an exponential. To be specific, consider
the term of order $g^2$ in the expansion (\ref{tilVpert}).
To $\tilde V({\vp}; u=-it)$, this terms contributes
($\tilde D({\bf q})\equiv \tilde D(0,{\bf q})$)
\beq
-g^2 T\int \frac{{\rm d}^3q}{(2\pi)^3}
\frac{\tilde D({\bf q})}{({\vq})^2}\,
\Bigl[n({\vp})-n({\bf v\cdot (p +q)})\,
{\rm e}^{-it({\vq})}\Bigr],\eeq
while to $  {\rm e}^{\beta({\vp})}
 \tilde V({\vp}; u=\beta-it)$
it contributes
\beq
-g^2 T\int \frac{{\rm d}^3q}{(2\pi)^3} 
\frac{\tilde D({\bf q})}{({\vq})^2}\,
{\rm e}^{\beta({\vp})}\,
\Bigl[n({\vp})-n({\bf v\cdot (p +q)})\,
{\rm e}^{\beta({\vq})}{\rm e}^{-it({\vq})}\Bigr].\eeq
In the sum of these two expressions,
the thermal factors disappear because of the identity (\ref{ID1}), leaving
\beq
-g^2 T\int \frac{{\rm d}^3q}{(2\pi)^3} 
\frac{\tilde D({\bf q})}{({\vq})^2}\,
\Bigl(1- {\rm e}^{-it({\vq})}\Bigr).\eeq
By analyzing similar compensations for the higher order terms,
we eventually recognize the power expansion of an exponential:
\beq\label{SRT}
S_R(t,{\bf p})&=&i\theta(t) {\rm e}^{-it({\vp})}\,\Delta(t),\eeq
with \beq\label{SR0}
\Delta(t)& \equiv & {\rm exp}\left \{-g^2T
\int \frac{{\rm d}^3q}{(2\pi)^3} 
\frac{\tilde D({\bf q})}{({\vq})^2}\,
\Bigl(1-  {\rm cos}\,t({\vq})\Bigr)\right\}.\eeq
The zero-frequency photon propagator reads
\beq\label{D0q}
\tilde D({\bf q})\,\equiv \,v^\mu\, {}^*D_{\mu\nu}(\omega_m=0, {\bf q})
v^\nu\,=\,-\frac{1}{q^2+m_D^2}+
\frac{1}{q^2}\left(1-\frac{({\vq})^2}{q^2}\right)
+\lambda \frac{({\vq})^2}{q^4},\eeq
in an arbitrary gauge of the Coulomb or the covariant type
($\lambda=0$ corresponds to both the Landau and the strict Coulomb gauges).
The three terms in eq.~(\ref{D0q}) corresponds respectively
to the electric, magnetic and gauge sector.
In eq.~(\ref{SR0}), we have replaced the complex
exponential by a cosine function, by taking into account
the parity of the integrand.

The  $q$-integral
in eq.~(\ref{SR0}) presents a spurious ultraviolet logarithmic divergence
in the physical sector (i.e., for electric and magnetic photons).
This divergence is unphysical since in the full theory,
including also the non-static photon modes, 
the $q$-integral would be cut-off at momenta
 $q\sim \omega_p$ (recall the discussion in
section 2.2).
Thus, to be consistent with the approximations performed,
we have to complement the above ``dimensional reduction'' 
with the prescription that an upper cut-off of the order $gT$
is added in momentum integrals, in the physical
 sector. Since this cut-off is
not exactly known, it will be important in what follows
to verify that  the physical predictions are independent
from its precise value. In the gauge sector, on the other hand,
 no such cut-off is needed since
the corresponding momentum integral turns out to be ultraviolet
finite (see below, eq.~(\ref{qlam})).

Eq.~(\ref{SR0}) determines the large time behavior of the fermion
propagator, to be discussed in the next section.
At a first sight, the considerable simplifications  leading
to this equation (and coming from the restriction to the static 
photon modes in eq.~(\ref{tilVpert})) may seem rather accidental.
However, as we explain now, there is a simple reason for
these simplifications, and, in fact,  eq.~(\ref{SR0}) could
have been obtained in a more direct way\cite{prl}, which avoids
some of the complications of the Matsubara formalism
(the latter  are essential only for the non-static modes).
Let us indeed return briefly to the functional integral
of eq.~(\ref{FS0}), and consider its approximation
where the Bloch-Nordsieck propagator $ G_E(x,y|A)$
includes only the {\it static} electromagnetic field $A_\mu({\bf x})$
of eq.~(\ref{Ast}). Then, the contribution of the non-static photon modes 
to the functional integral trivially factorizes, and is compensated by the
corresponding contribution to the partition function $Z$, thus leaving
\beq\label{FS}
S_E(x,y)= Z_0^{-1}\int [{\rm d} A]\, G_E(x,y|A)\,{\rm exp}\left\{-
\frac{1}{2}\,\Bigl(A,\, {}^*D^{-1}A\Bigr)_0\right\},\eeq
where  $A_\mu\equiv A_\mu(\omega_m=0,{\bf q})$, and
 $(A, {}^*D^{-1}A\Bigr)_0$ denotes the $\omega_m=0$ contribution to
 the effective photon action (\ref{SEFF}); correspondingly,
 $Z_0$ is the partition function of the static mode alone.
Since the  background  field (\ref{Ast}) is time-independent,
the propagator $ G_E(x,y|A)$ depends only on the time difference $x_0-y_0$, i.e.
$G_E(x,y|A)\equiv G_E(x_0-y_0, {\bf x, y}|A)$.
Its Fourier transform can be analytically continued
in the complex energy plane, and the resulting function coincides,
in the upper half plane, with the retarded propagator.
It is then convenient to take the Fourier transform of
eq.~(\ref{FS}), and write
($p^\mu=(i\omega_n,{\bf p})$, $\omega_n=(2n+1)\pi T$):
\beq\label{SEP} S_E(p)
\equiv Z_0^{-1}\int [{\rm d} A]
\,G_E({\bf x}, p|A) \,{\rm exp}\left\{-
\frac{1}{2}\,
\Bigl({A},\,{}^*D^{-1}{A}\Bigr)_0\right\}.\eeq
Since the energy $p_0$ enters eq.~(\ref{SEP}) as an external
parameter, the continuation
to real external energy $p_0\to \omega+i\eta$,
and the Fourier transform to real time, can both be performed
 {\it before} doing the functional integration. Thus,
the retarded propagator $S_R(t,{\bf x})$ can be directly obtained
 as the functional average of $G_R(x,y|A)$, which is known explicitly 
(recall eqs.~(\ref{GR}) and (\ref{U})).

 Specifically, eqs.~(\ref{SEP}) and (\ref{GR}) give
 $S_R(t,{\bf p})$ in the form (\ref{SRT}), where
\beq\label{Delta0}
\Delta(t)\equiv Z_0^{-1}\int [{\rm d} A]
\,U(x,x-vt) \,{\rm exp}\left\{-
\frac{1}{2}\,\Bigl(A, {}^*D^{-1}A\Bigr)_0\right\},\eeq
and the parallel transporter is that of a static background field:
\beq 
U(x,x-vt)&=&\exp\left\{ -\int {\rm d}^3 y\, j_\mu({\bf y}) A^\mu({\bf y})
\right\},\nonumber\\
 j_\mu({\bf y})&\equiv & ig v_\mu\int_0^t {\rm d}s \, 
\delta^{(3)}({\bf x-y-v}s).\eeq
A straightforward computation yields then
\beq\label{Delta1}
\Delta(t) &=&{\rm exp}\left \{\frac{1}{2}\,T\int
{\rm d}^3 x_1 {\rm d}^3 x_2\, j^\mu({\bf x_1})\, {}^*D_{\mu\nu}({\bf x_1}
-{\bf x_2})\, j^\nu({\bf x_2})\right \} \nonumber\\
&=& {\rm exp}\left\{-\frac{g^2}{2}\,T\int_0^t {\rm d}s_1 
\int_0^t {\rm d}s_2\, \tilde {\cal D}({\bf v}(s_1-s_2))\, 
\right \},\eeq
where
\beq
\tilde {\cal D}({\bf x})\equiv
\int \frac{{\rm d}^3q}{(2\pi)^3}\,
{\rm e}^{i{\bf q\cdot x}}\, \tilde D ({\bf q}) \eeq
is the Fourier transform of the static photon propagator
(\ref{D0q}). By using the last equation to perform the $s_1$ and $s_2$
integrations, we may cast eq.~(\ref{Delta1}) 
in the form (\ref{SR0}).

\section{The infrared structure of the fermion propagator}
\setcounter{equation}{0}
\subsection{Large time behavior}

The non-trivial time dependence of the fermion
propagator is contained in the function $\Delta(t)$, eq.~(\ref{SR0}).
Because our approximations preserve only
 the leading infrared behavior of the perturbation theory,
eq.~(\ref{SR0}) describes  only the leading {\it large-time} behavior
of $\Delta(t)$. Since the only energy scale in the momentum integral of
eq.~(\ref{SR0}) is the upper cut-off, of order $gT$,
 the large-time regime is achieved for $t\gg 1/gT$.

The expansion of  eq.~(\ref{SR0}) in powers of $g^2$ reproduces
the dominant singularities of the usual perturbative expansion
for the self-energy.
Let us verify this for the correction of order $g^2$:
\beq
\delta S_R(\omega,{\bf p}) = -g^2T\,
i\int_0^{\infty}{\rm d}t\,
{\rm e}^{it(\omega- {\vp} +i\eta)}\,
\int \frac{{\rm d}^3q}{(2\pi)^3} 
\frac{\tilde D({\bf q})}{({\vq})^2}\,
\Bigl(1- {\rm cos}\,t({\vq})\Bigr).\eeq
We perform first the time integration
and obtain, after simple algebraic manipulations,\footnote{
The self-energy $\Sigma$ which appears here 
corresponds to the spin projection $\Sigma_+$ of the full
self-energy. (See eq.~(\ref{PIpm}).)}
\beq
\delta S_R(\omega,{\bf p}) = - G_0(\omega,{\bf p}) \Sigma(\omega,{\bf p})\,
G_0 (\omega,{\bf p}),\eeq
where \beq\label{G0R}G_0(\omega, {\bf p})
\,=\,i\int_0^{\infty}{\rm d}t
\,{\rm e}^{it(\omega- {\vp}+i\eta)}\,=\,\frac{-1}{\omega-{\bf v\cdot p} +i\eta}
\,,\eeq is the free BN propagator, and 
 \beq\label{SigBN}
 \Sigma(\omega,{\bf p})&=&-g^2T \int \frac{{\rm d}^3q}{(2\pi)^3} 
\,\tilde D({\bf q})\,\frac{-1}{\omega - {\bf v}\cdot ({\bf p+q})
 + i\eta}\,.\eeq
The imaginary part of this equation  determines the damping rate according to
$\gamma = -  {\rm Im}\,  \Sigma(\omega=p)$. We can write, with ${\epsilon}\equiv \omega -{\vp}$,
\beq\label{ImSig}  {\rm Im}\,  \Sigma(\omega, {\bf p}) &=&
-\pi g^2T \int \frac{{\rm d}^3q}{(2\pi)^3} 
\,\delta(\omega - {\bf v}\cdot ({\bf p+q}))\,\tilde D({\bf q})\nonumber\\
&=&-\frac{g^2T}{4\pi}\,\int_{|\epsilon|}^{\omega_p}{\rm d}q\,q
\left\{
\frac{1}{q^2}\left(1-\frac{\epsilon^2}{q^2}\right) -\frac{1}{q^2+m_D^2}
+\lambda\, \frac{\epsilon^2}{q^4}\right\},\eeq
which, in the mass-shell limit $\epsilon \to 0$, and with an IR cut-off
$\mu$ in the magnetic sector, yields (with $\alpha =g^2/4\pi$)
\beq\label{gammaBN}
\gamma= \alpha T \left\{\ln \frac{\omega_p}{\mu} \,-\,\frac{1}{2}\ln\left(
1+\frac{\omega_p^2}{m_D^2}\right)\,+\,\frac{\lambda-1}{2}\right\}\,.\eeq
This first piece inside the parantheses, which comes from the magnetic sector,
 reproduces the singular piece of the resummed one-loop calculation (recall
eq.~(\ref{gammap1})). On the other hand, the other two pieces are not correctly
reproduced by the present calculation. The electric piece, which is finite
and of the order $g^2T$, occurs even with a minus sign
(recall that the contribution of the electric scattering
to the interaction rate in eq.~(\ref{g5}) was {\it positive}).
The gauge-dependent piece turns out to be non-vanishing,
but it could be eliminated by introducing an IR cut-off $\mu$ in the gauge
sector as well, and by taking the on-shell limit only subsequently \cite{Rebhan93}
(see also the discussion in Appendix B).
This situation is generic: our approximation yields correctly only
the leading IR divergences of the usual perturbation
theory, which all arise from the magnetic sector, but not
 the subleading terms. In particular, the contributions
involving the electric and the gauge sector are 
subleading, and {\it should be discarded for consistency.}
This is equivalent to using $\tilde D({\bf q}) = 1/q^2$ rather
than the full static propagator of eq.~(\ref{D0q}).

Let us verify now that the full, non-perturbative, expression
of $\Delta(t)$, eq.~(\ref{SR0}), is {\it free of infrared singularities}.
Inspection of the integrand in eq.~(\ref{SR0}) shows that the dominant
IR behaviour arises from the limit $|{\vq}|\equiv q\cos \theta \,\to 0$.
This is consistent with the calculations in sections 2.1 -- 2.2
showing that the divergences come from the exchange of magnetic photons
emitted or absorbed at nearly 90 degrees. We have, in this limit,
\beq\label{approx}
\frac{1-  {\rm cos}\,t({\vq})}
{({\vq})^2} \,\simeq\,\frac{t^2}{2} + {\cal O}\Bigl(t^4 ({\vq})^2\Bigr),\eeq
and the momentum integral is IR safe, as advertised.
We see here, once again, that the gauge dependent piece
of the photon propagator (\ref{D0q}) does not contribute
to the leading IR behavior (which is given by the term in
$1/q^2$ of the magnetic propagator).
 Indeed,  because of the factor $(\cos\theta)^2$,
the gauge propagator $(\cos\theta)^2/q^2$ is less singular as $q\cos \theta \to 0$.

Consider now the UV behavior of the $q$-integral.
This depends logarithmically on the UV cut-off
$\sim \omega_p$, and, as a consequence,
the large time behavior of
$\Delta(t)$ is insensitive  to both the precise
value  of the UV cut-off, and to the specific procedure which is
used for its  implementation. This will be verified explicitly below.

The evaluation of $\Delta(t)$ is most simply
done by using the coordinate space representation
(\ref{Delta1}) for $\Delta(t)$. Corresponding 
to $\tilde D(q)=1/q^2$, we have
$\tilde {\cal D}({\bf x})\,=\,
 1/4\pi x$, and we obtain, for $t \gg 1/\omega_p$,
$\Delta(t)={\rm exp}(-g^2T\,F(t))$, with
\beq\label{qint}
F(t)&\equiv& \frac{1}{2}\,\int_0^t {\rm d}s_1 
\int_0^t {\rm d}s_2\, \tilde {\cal D}({\bf v}(s_1-s_2))
\nonumber\\&=&
\frac{1}{8\pi}\int_0^t {\rm d}s_1 
\int_0^t {\rm d}s_2\, \frac{\theta(|s_1-s_2|-1/\omega_p)}{|s_1-s_2|}
\simeq\,\frac{t}{4\pi}\left(\ln \omega_pt \,+{\cal O}(1)\right ).\eeq
In this calculation, the  ultraviolet cut-off
has been introduced in the function
 $\theta(|s_1-s_2|-1/\omega_p)$.
Let us verify that the same large time behavior
is obtained with a different UV regularisation,
namely, with the modified photon propagator
 $\tilde D(q)=1/q^2 - 1/(q^2+\omega_p^2)$
(Pauli-Villars regularisation). By using
\beq 
\tilde {\cal D}({\bf x})=\int \frac{{\rm d}^3q}{(2\pi)^3}\,
{\rm e}^{i{\bf q\cdot x}} \left (\frac{1}{q^2}
-\frac{1}{q^2+\omega_p^2}\right)=\frac{1}{4\pi x}\Bigl
(1-{\rm e}^{-\omega_p x}\Bigr),\eeq
we get successively
\beq\label{qint1}
F(t)&=&\frac{1}{8\pi}\int_0^t {\rm d}s_1 
\int_0^t {\rm d}s_2 \frac{1-{\rm e}^{-\omega_p|s_1-s_2|}}{|s_1-s_2|}
\nonumber\\&=&
\frac{t}{4\pi}\left\{\frac{1-{\rm e}^{-\omega_p t}}{\omega_pt}\,-1\,+
\int_0^{\omega_pt} {\rm d}s\,\frac{1-{\rm e}^{-s}}{s}\right\}\nonumber\\
&=&\frac{t}{4\pi}\left\{\ln \omega_pt + (\gamma_E-1) +\frac
{1-{\rm exp}(-\omega_pt)}{ \omega_pt} + {\rm E}_1 (\omega_pt)\right\},\eeq
where ${\rm E}_1(x) $ is the exponential-integral function,
 ${\rm E}_1(x)=\int_1^\infty {\rm d}y \,({\rm e}^{-xy}/y)$,
and $\gamma_E$ the Euler constant.
At very large times, $\omega_pt\gg 1$, we may use the asymptotic expansion
of  the exponential-integral to get, for the r.h.s. 
of eq.~(\ref{qint1}),
\beq\label{qLT}
F(t)\simeq (t/4\pi)\Bigl(\ln \omega_pt + (\gamma_E-1)\Bigr)\simeq
(t/4\pi)\ln \omega_pt,\eeq
which coincides, as long as the leading logarithm
is concerned, with the previous result (\ref{qint}).
On the other hand, the subleading term, i.e. the constant under 
the logarithm, is dependent on the UV regularisation.
Thus, as expected, it is only the dominant behavior
at very large times which is consistently described
by our approximation; the subleading terms should be ignored.

We have argued before that the gauge-fixing terms are not
important to the order of interest. To verify this explicitly,
 we compute the gauge-dependent contribution
to $F(t)$, as given by the last term of the photon propagator
(\ref{D0q}):
\beq\label{qlam}
\delta F(t)\equiv \lambda\int \frac{{\rm d}^3q}{(2\pi)^3} 
\,\frac{({\vq})^2}{q^4}\,
\frac{1-  {\rm cos}\,t({\vq})} {({\vq})^2}\,=\,
\lambda\,\frac{t}{8\pi}.\eeq
At large times, this is indeed subleading with respect to (\ref{qLT}).
Note that, the momentum integral in eq.~(\ref{qlam}) being
ultraviolet finite,  no upper cut-off has been necessary
in its evaluation.

We conclude that, at times $t\gg 1/\omega_p$,
the function $\Delta(t)$ is gauge-independent and of
the form  ($\alpha =g^2/4\pi$)
\beq\label{DLT}
\Delta(\omega_pt\gg 1)\simeq {\rm exp}\Bigl( -\alpha Tt \ln \omega_p
t\Bigr).\eeq
The most striking feature of this  result
is the fact that, at very large times ($\omega_p t\to \infty$), 
the fermion propagator is decreasing faster than any exponential. 
We also note that the scale of the time variations
is fixed by the plasma frequency $\omega_p\sim gT$.

\begin{figure}
\protect \epsfysize=11.5cm{\centerline{\epsfbox{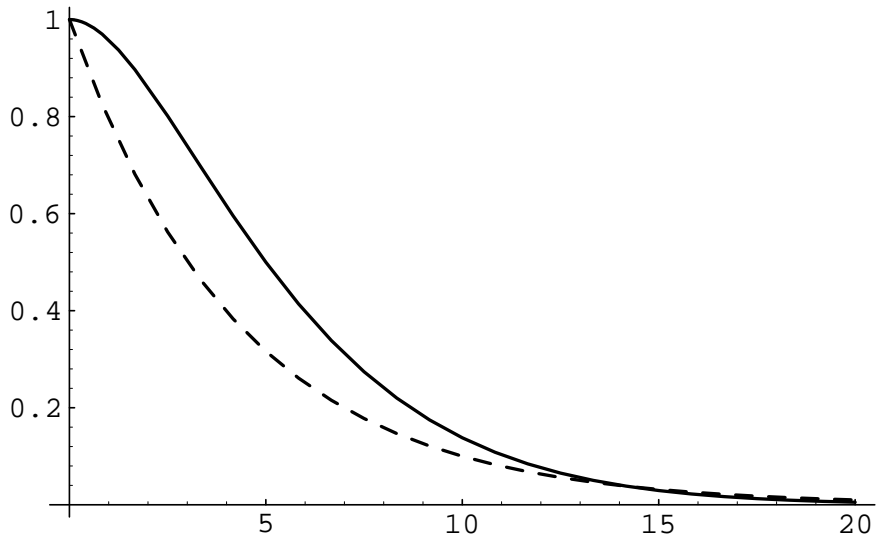}}}
	 \caption{The time behavior of the fermion propagator
as described by the nonperturbative result $\Delta(t)$ (full line)
and by the exponential $\Delta_L(t)$ (dashed line) for $g=0.4$. On the abscissa
axis, time is measured in units of $1/\omega_p$.}
\label{delfig}
\end{figure}

A measure of the decay time $\tau$ is given by
\beq\label{tauex} \frac{1}{\tau}=\alpha T\ln \omega_p \tau=
\alpha T\left(\ln \frac{\omega_p}{\alpha T} - \ln
\ln \frac{\omega_p}{\alpha T} + \,...\right).\eeq
Since $\alpha T \sim g\omega_p$, we see that $\tau \sim
 1/(g^2 T \ln (1/g))$. This is very close to
the perturbative result (\ref{gammap1}), which, in the presence of an IR cut-off
$\sim g^2T$, predicts a damping rate $\gamma \sim g^2 T \ln (1/g)$.
A comparison of the two decay laws, $\Delta(t)={\rm exp}(-g^2T\,F(t))$, with
$F(t)$ from eq.~(\ref{qint1}), and the exponential\footnote{
This is the spectral function which would produce an exponential
decay in time with a lifetime as close as possible to the
nonperturbative result (\ref{tauex}).}
$\Delta_L(t)=\exp(-\gamma t)$ with $\gamma= \alpha T \ln(1/g)$,
is presented in Fig.~\ref{delfig} for $g=0.4$.
In this figure, the time is measured in units of $1/\omega_p$, and
the results  displayed for $\Delta(t)$ can be trusted for values $\omega_p t>>1$,
where our approximations are expected to hold. 
For very large times, $t\gg \tau$, the function $\Delta(t)$
is indeed more rapidly decreasing than the exponential $\Delta_L(t)$.
 On the other hand, for intermediate, but still large,
 times, $1/gT \ll t \ll 1/g^2T$, the opposite
situation holds: $\Delta(t) > \Delta_L(t)$.
 When discussing the lifetime of the excitation, it is rather the
intermediate range of times which matters, since for
asymptotically large times $t\simge 1/g^2T$ the excitation
has already decayed. It follows that, for the range of times
of interest, the decay of the excitation is actually {\it slower} than
the one predicted by perturbation theory.

\subsection{Mass-shell behavior}

The non-trivial large-time behavior exhibited in eq.~(\ref{DLT})
 has interesting consequences on the
 behavior of the retarded propagator in the complex energy plane.
In fact, since at large times $\Delta(t)$
is decreasing faster than any exponential,
 the time-integral giving the Fourier transform
\beq\label{SRO}
S_R(\omega, {\bf p})\,=\,
\int_{-\infty}^{\infty} {\rm d}t \,{\rm e}^{-i\omega t}
S_R(t,{\bf p})\,=\,
i\int_0^{\infty}{\rm d}t
\,{\rm e}^{it(\omega- {\bf v\cdot p}+i\eta)}\,\Delta(t),\eeq
 is absolutely convergent
for {\it any} complex (and finite) $\omega$. That is, the retarded propagator
 $S_R(\omega)$ is an entire function, with sole singularity
at ${\rm Im}\,\omega\to -\infty$. Recall, however, that strictly
 speaking, our present approximation holds only
in the vicinity of the mass-shell. Therefore, when speaking
about $|\omega-{\vp}|\to \infty$ we have in mind off-shell
deviations which are much larger than $g^2T$.
To further clarify this point, let us give a crude estimate
of how $S_R(\omega)$ increases as 
${\rm Im}\,\omega\to -\infty$. To this aim, let us consider
 $\omega={\bf v\cdot p} - i\zeta$, with
real and positive $\zeta$. We write:
\beq\label{Szeta}
S_R(\zeta)\,=\,
i\int_0^{\infty}{\rm d}t
\,{\rm e}^{\,\zeta t}\,\Delta(t),\eeq
which is a purely imaginary function of $\zeta$,
and consider the behaviour of $|S_R(\zeta)|$ for $\zeta \gg \alpha T$.
Regarded as a function of $t$, the integrand ${\rm e}^{\,\zeta t}\,\Delta(t)$
is rapidly increasing for small $t$, but it is decreasing for
sufficiently large values of $t$, where the decay of $\Delta(t)$
starts to dominate. Assuming the time integral in (\ref{Szeta}) to be dominated
by large values of $t$,  --- which is 
correct for large enough $\zeta$ ---, we can use the asymptotic
expression (\ref{DLT}), and determine the time $t^*$ at which
the integrand is maximum:
\beq\label{t*}
\omega_p t^*\,=\,\exp{\frac{\zeta - \alpha T}{\alpha T}}\,.\eeq
By using the fact that the integrand is  positive definite, and that,
according to eq.~(\ref{t*}), $\zeta- \alpha T \ln \omega_pt^*=  \alpha T$
and thus $\zeta- \alpha T \ln \omega_pt >  \alpha T$ for any $t < t^*$,
we can write\footnote{This estimate was suggested to us
 by A. Rebhan \cite{pers}}
\beq
|S_R(\zeta)|&>& \int_0^{t^*}{\rm d}t
\,{\rm e}^{\,\alpha T t}\,=\,\frac{1}{\alpha T}\,\Bigl(
{\rm exp}(\alpha T t^*)-1
\Bigr),\eeq so that
\beq\label{lower}
|S_R(\zeta)|\,>\,\frac{1}{\alpha T}\,\left\{{\rm exp}\left(\tilde g\,
{\rm exp}{\frac{\zeta - \alpha T}{\alpha T}}\right)\,-\,1\right\}\,\simeq\,
\,
\frac{1}{\alpha T}\,{\rm exp}\left(\tilde g\,
{\rm exp}{\frac{\zeta}{\alpha T}}\right)\,,\eeq
where $\tilde g \equiv \alpha T/\omega_p = (3/4 \pi)\,g$.
Eq.~(\ref{lower}) shows that $|S_R(\zeta)|$ is rapidly
increasing starting with values of $\zeta$ such that
$\tilde g
{\rm e}^{\frac{\zeta}{\alpha T}} \sim 1$, that is,
$\zeta \sim {\alpha T} \ln(1/g)$. In perturbation theory,
$S_R(\omega)$ has a pole at $\omega={\bf v\cdot p} - i\gamma$, where
$\gamma \simeq \alpha T \ln(\omega_p/\mu)
\simeq \alpha T \ln(1/g)$ if
$\mu \sim g^2T$. Thus, our non-perturbative solution
for $S_R(\omega)$ replaces the pole at finite distance 
by an essential singularity at $-i\infty$, which however starts manifesting 
itself at distances $\sim g^2 T\ln (1/g)$ below the real axis,
that is,  at the same distances as the pole of the perturbation theory.

\begin{figure}
\protect \epsfxsize=10.cm{\centerline{\epsfbox{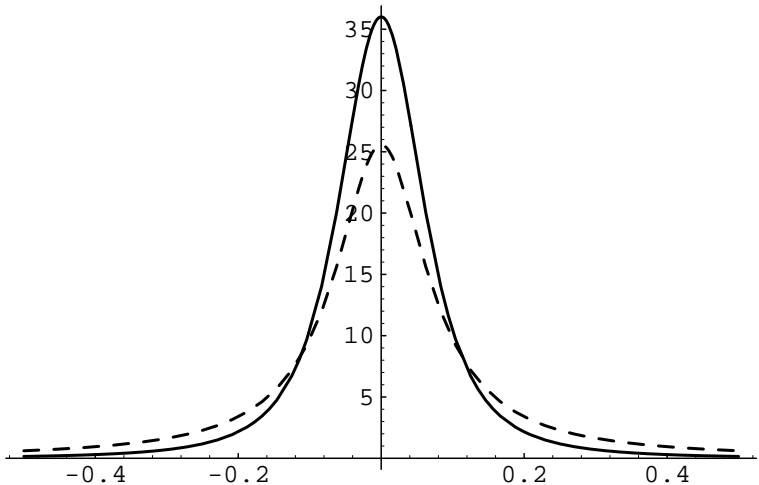}}}
	 \caption{The spectral density $\acute\rho(\epsilon)$ (full line, in units
of $1/\omega_p$)
and the lorentzian $\rho_{L}(\epsilon)$ (dashed line) for $g=0.08$,
as a function of  $\epsilon\equiv v \cdot p$ in units of $\omega_p$.}
\label{ac2}
\end{figure}

Since  $S_R(\omega)$ is analytic in any finite
neighbourhood of the tree-level mass-shell at $\omega={\vp}$, we need to
 clarify the mass-shell interpretation. To this aim, we
consider the spectral density $\acute\rho(\omega, {\bf p})$
\beq\label{rhoD}
\acute\rho(\omega, {\bf p})= 2\,{\rm Im}\,S_R(\omega, {\bf p})
= 2 \int_0^{\infty}{\rm d}t\,\cos\,t(v\cdot p)\,\Delta(t),\eeq
where $v\cdot p= \omega -{\vp}$.
It   satisfies the sum-rule\footnote{In fact, this sum rule 
holds exactly in the Bloch-Nordsieck
model, independently of the restriction to the static
photon mode. In general, $\Delta(t=0)$ is replaced, in eq.~(\ref{SRule}),
by $S^>(t=0, {\bf p})+S^<(t=0, {\bf p})$, which is also equal
to one, as shown by eq.~(\ref{sum}).}
\beq\label{SRule}
\int_{-\infty}^\infty {{\rm d}\omega \over 2\pi}\,
\acute\rho(\omega, {\bf p})=\Delta(t=0)=1.\eeq
We have calculated $\acute\rho(\omega, {\bf p})$ numerically,
and the result is plotted, for a coupling constant $g=0.08$, in Fig.~\ref{ac2}.
We also represent, for the same value of $g$, the Lorentzian spectral function
($\epsilon\equiv v \cdot p$)
\beq\label{lor}
\rho_{L}(\epsilon)\,=\,\frac{2\gamma}{\epsilon^2 + \gamma^2}\,,\eeq
with $\gamma= \alpha T \ln(1/g)$.
This is the spectral function which would produce the exponential
time decay  $\Delta_L(t)=\exp(-\gamma t)$
alluded to at the end of the previous subsection. It is seen
on these figures that, in the weak coupling limit,
 the spectral density $\acute \rho(\epsilon)$
 has the shape of a {\it resonance}  strongly peaked around $\epsilon =0$,
and with a typical witdh of the order $1/\tau\sim g^2T \ln(1/g)$,
that is, of the same order as that of the Lorentzian.
This allows us to identify the mass-shell of the full propagator
at  $\omega={\vp}$, as at treel-level.
Moreover, it is clear from Fig.~\ref{ac2} that,
 for very small $g\ll 1$, the nonperturbative
spectral density is even {\it sharper}  than a Lorentzian.
Thus the net result of the infrared effects considered here is to slightly
{\it enhance} the stability of the quasiparticle state
(see also the discussion at the end of the previous subsection).

\begin{figure}
\protect \epsfysize=13.cm{\centerline{\epsfbox{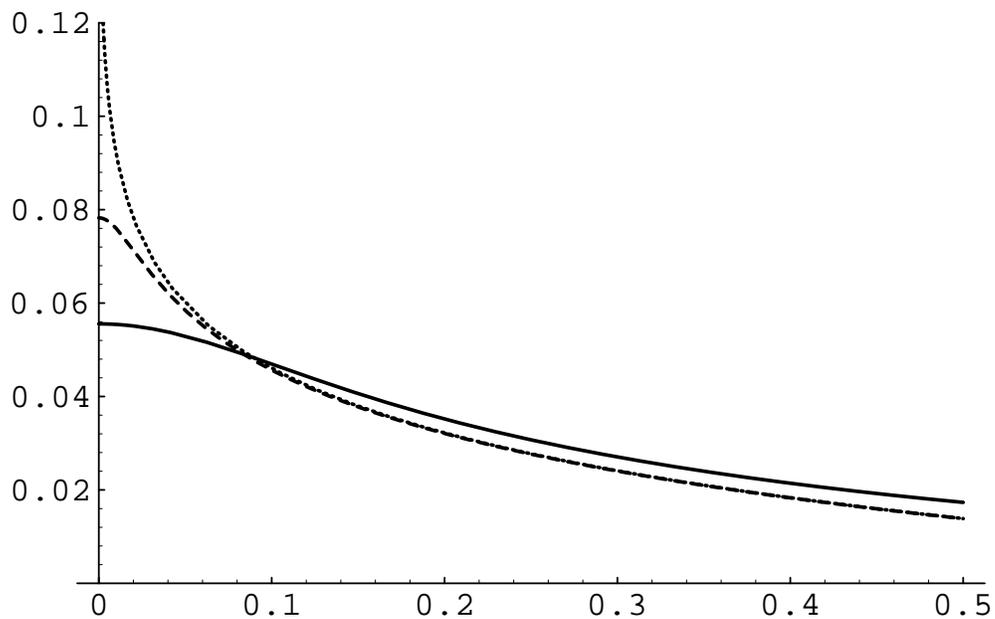}}}
	 \caption{The imaginary part of the self-energy, eq.~(\ref{ImSig1}),
as a function of the energy, for $g=0.08\,$:
nonperturbative calculation (full line), one-loop result (dotted line)
and one-loop result in the presence of an IR cut-off $\sim g^2 T$ (dashed line).
All the quantities are measured in units of $\omega_p$.}
\label{IMS}
\end{figure}

Finally, it is interesting to compute the imaginary part of the
exact self-energy, by inverting the Dyson-Schwinger equation
$S_R^{-1}(\omega, {\bf p})=-(\omega - {\bf v\cdot p}) + \Sigma_R
(\omega, {\bf p})$. A simple calculation yields
\beq\label{ImSig1}
-{\rm Im}\Sigma_R(\epsilon)\,=\,\frac{2 \acute\rho(\epsilon)}
{\acute\sigma^2(\epsilon) +\acute\rho^2(\epsilon) }\,,\eeq
where  $\epsilon\equiv v \cdot p$, $\acute\rho(\epsilon)$ is the spectral
density of eq.~(\ref{rhoD}) and
\beq\label{sigmaD}
\acute\sigma(\epsilon)\equiv 2 \int_0^{\infty}{\rm d}t
\,\sin \epsilon t\,\Delta(t).\eeq
This is represented graphically in Fig.~\ref{IMS},
together with the pure one-loop result, eq.~(\ref{ImSig}),
which shows a logarithmic divergence as $\epsilon \to 0$ (dotted line),
and the screened one-loop result, as obtained from eq.~(\ref{SigBN})
after inserting an IR cut-off equal to $\alpha T$ (dashed line).
As manifest on this figure,
 the full result for ${\rm Im}\,\Sigma_R$ is finite
at the mass-shell $\epsilon=0$, and inferior to the value predicted
by the perturbation theory with an IR cut-off $\sim g^2T$. 
The latter property is consistent with the
the previous analysis of the spectral density, and also
of the time behavior at intermediate times.
One can also verify the non perturbative character of the solution.
For example,
${\rm Im}\,\Sigma_R(\epsilon=0)=-1/\int_0^{\infty}{\rm d}t \,\Delta(t)$
has no expansion in powers of $g^2$ even if one keeps
$\omega_p$ constant in eq.~(\ref{DLT}) for $\Delta(t)$.

\setcounter{equation}{0}
\section{The lifetime of the soft fermionic excitations}

For soft momenta, $p\sim gT$, the quasiparticles become
collective excitations, with non-trivial 
dispersion relations\cite{Klimov81,Weldon82}
and self-interactions\cite{BP90,FT90}.
 To leading order in $g$, the dispersion relations are real, and the
quasiparticles propagate without damping.
At next to leading order, collisional damping occurs.
The corresponding damping rate $\gamma$ has been calculated
in the effective (i.e., HTL-resummed) perturbation theory\cite{BP90}.
For an excitation with zero momentum ($p=0$), $\gamma$ is finite
and of the order $g^2T$ \cite{BP90,KKM}. However, for excitations
with finite momentum $p\gg g^2T$, the lowest order perturbative
calculation of $\gamma$ meets with the same
infrared problem as that discussed
 for the hard particles\cite{Pisarski93,Rebhan95}.
As we shall see, this problem is solved by the same technique
as that used for the hard fermion.

\subsection{The HTL approximation}

Let us recall first the main features of
 the dispersion relations for soft fermions,
to leading order in $g$. They are obtained from the poles
of the effective propagator ${}^*S(\omega, {\bf p})$
which is obtained as ${}^*S^{-1}=S^{-1}_0+\delta\Sigma$,
with $\delta \Sigma(\omega, {\bf p})$ denoting the
fermion self-energy in the HTL approximation\cite{Klimov81,Weldon82}:
\beq\label{Sigma}
\delta
\Sigma(\omega, {\bf p})
 =\omega_0^2\int\frac{{\rm d}\Omega}{4\pi}\, \frac{\slashchar{v}}
{\omega- {\bf v}\cdot {\bf p}+i\eta}\,.\eeq
In this equation,
$\omega_0 =gT/\sqrt{8}$ is the frequency of the spatially uniform ($p=0$)
fermionic excitations.
The propagator is conveniently written in the form (\ref{Sfull}),
that is,
\beq\label{Ssoft}
{}^*S(\omega, {\bf p})={}^*\Delta_+(\omega,p) h_+(\hat {\bf p})
+{}^*\Delta_-(\omega,p) h_-(\hat {\bf p}),\eeq
where
\beq
{}^*\Delta_\pm(\omega,p)\,=\,
\frac{-1}{\omega\mp (p + \delta\Sigma_\pm(\omega,p))}\,,\eeq
and 
\beq
 \delta\Sigma_\pm (\omega,p) = \pm \,\frac{1}{2}\,{\rm tr}\,\Bigl(
h_\pm(\hat {\bf p}) \delta\Sigma(\omega, {\bf p})\Bigr).\eeq
The pole equations ${}^*\Delta^{-1}_\pm(\omega(p),p)=0$
yield two positive energy branches $\omega_\pm (p)$
\cite{Klimov81}, instead of the usual one (with $\omega=p$)
in the free electron spectrum. For $\omega$ close to the mass-shell
at $\omega_s(p)$, $s=\pm$, we can write
\beq\label{res0}
{}^*\Delta_s(\omega,{\bf p})&\simeq&
\frac{z_s(p)}{\omega_{s}(p)-\omega}\,,\eeq
where  $z_s(p)$ is the residue of the mode $s\,$,
\beq\label{vpm}
z_{s}^{-1}(p)&=& 1-\frac {\del\, \delta\Sigma_s (\omega,p)}
{\del\omega}\Bigl|_{\omega=\omega_s(p)}\,.\eeq
Since $\omega_\pm (p)> p$ for any $p$, both dispersion relations
are real: the quasiparticles propagate without damping
in this approximation. 
For small momenta, $p\ll gT$,
$\omega_{\pm}(p)\simeq \omega_0 \pm p/3$.
The upper branch is strictly increasing
($v_+(p)>0$ for any $p$), while the lower branch
has a minimum at $p=p_c\approx 0.92\, \omega_0$. At very large momenta,
 $p\gg \omega_0$, both branches approach the light cone,
but $z_+(p)\to 1$, while $z_-(p)$ vanish exponentially.
(See Refs. \cite{BOllie,BIO96,MLB96}
for more details and physical interpretation.)

Because of the  gauge symmetry, the nonlocal character
of the HTL self-energy (\ref{Sigma}) leads to
effective interactions between a fermion pair and any number
of soft photons. For instance, the Ward identity
\beq\label{wid}
q^\mu\,{}^*\Gamma_\mu(p, p+q)= {}^*S^{-1}(p)- {}^*S^{-1}(p+q)\,,
\eeq
requires the existence of a nonlocal 3-point vertex function,
which is indeed found in the form
${}^*\Gamma_{\mu}(p,p+q)= \gamma_\mu + \delta\Gamma_{\mu}(p,p+q)$,
where $\delta\Gamma_{\mu}(p,p+q)$ is the 3-point HTL \cite{BP90,FT90}:
\beq\label{3gamma}
\delta\Gamma_{\mu}(p,p+q)\,=\,
\omega_0^2 \int\frac{{\rm d}\Omega}{4\pi}\frac{v_\mu
{\slashchar v}}{(v\cdot p+i\eta)(v\cdot (p+q)+i\eta)}.\eeq
Similarly, higher vertices, without analogue at the
tree-level, are necessary in order to fulfill the
higher order Ward identities. We show here one more example,
namely the Ward identity satisfied by the  2-fermions --- 2-photons
vertex function:
\beq\label{wid2}
q_1^\mu\,{}^*\Gamma_{\mu\nu}(p_1,p_2;q_1,q_2)= {}^*\Gamma_\nu(p_1,p_1+q_2)
- {}^*\Gamma_\nu(p_1+q_1,p_1+q_1+q_2)\,,\eeq
where, in the left hand side, $p_i$ and $q_i$ are respectively
the momenta of the incoming fermions and photons,
with $p_1+p_2+q_1+q_2=0$.
For what follows, it is important to remark that all the 
HTL vertex functions are  (almost) uniquely determined by the self-energy
(\ref{Sigma}) and the Ward identities like eq.~(\ref{wid})
\cite{FT90,BP90b}. This is so since the non-linear
structure of the effective action of the HTL's
is the minimal one which is 
consistent with the gauge symmetry\cite{FT90,FT92,qcd}.

\subsection{Perturbation theory for the damping rate}

\begin{figure}
\protect \epsfxsize=11.cm{\centerline{\epsfbox{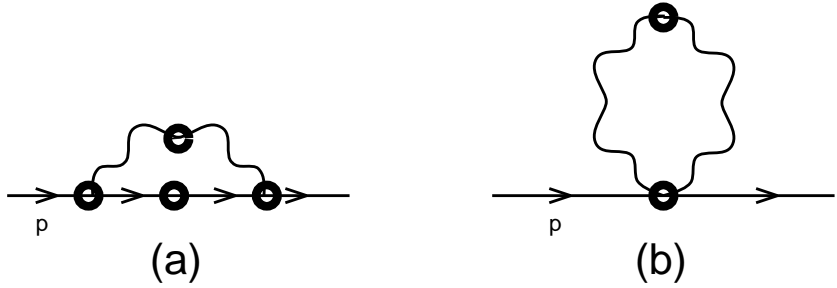}}}
	 \caption{One-loop diagrams for the soft fermion self-energy in the
effective expansion.}
\label{figeff}
\end{figure}

In this section, we discuss the perturbative computation
of the damping rate for the soft fermion, and the related IR problems.
 After a brief summary of the leading-order 
computation \cite{BP90,Pisarski93,Rebhan95}, we discuss
higher orders corrections and how they simplify in the computation
of the leading divergent terms.

The dominant contribution to the damping
rate, of order $g^2T$, comes from the imaginary part
 of the (resummed) one-loop
self-energy ${}^*\Sigma(\omega, {\bf p})$,
as given by the two diagrams in Fig.~\ref{figeff} \cite{BP90}.
Specifically, 
\beq\label{gpm} \gamma_\pm(p)
\,=\,- z_{\pm}(p)\,{\rm Im}\,{}^*\Sigma_\pm (\omega_\pm(p)+i\eta, p),\eeq
where ${}^*\Sigma_\pm(p)={\rm tr}\,(h_\pm(\hat {\bf p}) {}^*\Sigma(p))/2$
and the subscripts $\pm$ refer to the two positive-energy modes
in the fermion spectrum.
Note that, in general, the ``tadpole'' diagram in
Fig.~\ref{figeff}.b gives a non-trivial contribution 
to $\gamma$,  since
the 4-point vertex itself has a non-zero discontinuity. Moreover,
the imaginary part of the diagram in Fig.~\ref{figeff}.a 
comes not only from the cutting of the internal propagators
(as for the usual one-loop diagram discussed in section 2),
but also from the discontinuity of the resummed 3-point vertices.

In what follows, we concentrate
 on the singular contribution to $\gamma$. This
 comes entirely from the diagram in Fig.~\ref{figeff}.a
\cite{Pisarski93}, which reads
\beq\label{Sigstar}
{}^*\Sigma_a(p)
=-\,g^2 T\sum_{q^0= i\omega_m}
 \int \frac{{\rm d}^3q}{(2\pi)^3}\,
\,{}^*\Gamma_\mu (p,p+q)\,{}^*S(p+q)\,{}^*\Gamma_\nu (p+q,p)
\,{}^*D^{\mu\nu}(q)\,.\eeq
It has been already recognized\cite{Pisarski93} that the singular
piece of $\gamma$ arises from the same kinematical regime
as for a hard fermion, namely, from the exchange of a very
soft ($q\simle g^2T$) magnetic photon 
at nearly 90 degrees. This allows for kinematical approximations
identical to those encountered in section 2. In particular, the whole 
singularity can be reproduced by  restricting the calculation
to the {\it  magnetostatic} mode $q^0= i\omega_m=0$ \cite{Rebhan95},
with propagator ${}^*D^{ij}(\omega_m=0,{\bf q})=\delta^{ij}/q^2$,
and with an upper cut-off $\sim gT$.
Furthermore, the internal fermion propagator
${}^*S(\omega, {\bf p+q})$ is  nearly on-shell, since
 $\omega\simeq \omega_\pm(p)$, and $q\ll p$. 
Thus we can write $\omega_{\pm}({\bf p+q})\simeq \omega_{\pm}(p)
+{\bf v}_{\pm}(p)\cdot {\bf q},$ ---
where ${\bf v}_{s}(p)$ denotes the  group velocity of the mode $s$,
${\bf v}_{s}(p) \equiv \del \omega_{s}(p)/\del {\bf p} =
  v_s(p)\,\hat {\bf p}$ ---,
and replace ${}^*S(\omega, {\bf p+q})\longrightarrow
{}^*\Delta_{\pm}(\omega,{\bf p+q})\,h_\pm(\hat {\bf p})$,
with (recall eq.~(\ref{res0}))
\beq\label{res}
{}^*\Delta_{\pm}(\omega,{\bf p+q})&\simeq&
\frac{-z_{\pm}(p)}{\omega-\omega_{\pm}(p)
-{\bf v}_{\pm}(p)\cdot {\bf q}}\,,\eeq
where the upper (lower) sign applies
according to whether the external line is close to the mass-shell
of the upper branch, or of the lower branch, respectively.
A final simplification refers to the  3-point vertex 
function ${}^*\Gamma^i({\bf p,p+q})$, where we can neglect
the photon momentum ${\bf q}$ and use the differential form of the
Ward identity (\ref{wid}) to write:
\beq\label{diffW}
{}^*\Gamma^i({\bf p,p})\,=\,\frac{\del\, {}^*S^{-1}(\omega,{\bf p})}
{\del p^i}\,.\eeq
(The dependence of the vertex function on the external energy
$\omega$ is not indicated explicitly.)
The inverse propagator is conveniently written as (recall
eq.~(\ref{Sfull}))
\beq\label{Sinv}
{}^*S^{-1}(\omega, {\bf p})={}^*\Delta_+^{-1}(\omega,p) h_-(\hat {\bf p})
+{}^*\Delta_-^{-1}(\omega,p) h_+(\hat {\bf p}).\eeq
From eqs.~(\ref{diffW}) and (\ref{Sfull}), we obtain, for
$\omega\simeq \omega_\pm(p)$,
\beq \label{Gproj}
h_\pm(\hat {\bf p}){}^*\Gamma^i ({\bf p,p}) h_\pm(\hat {\bf p})&=&
h_\pm(\hat {\bf p})\,\frac{\del\, {}^*\Delta_{\pm}^{-1}}{\del p^i}
\,\simeq\,h_\pm(\hat {\bf p})\,\frac{v^i_\pm(p)}{z_\pm(p)}\,
\nonumber\\
{\rm tr}\,\Bigl(h_\pm(\hat {\bf p}){}^*\Gamma^i ({\bf p,p})
\Bigr)&=&2\,
\frac{\del\,{}^*\Delta_{\pm}^{-1}}{\del p^i}\,\simeq\, 
2\,\frac{v^i_\pm(p)}{z_\pm(p)}\,
.\eeq
The particular spin projections of ${}^*\Gamma^i$ written down
above are the only ones which enter 
 ${}^*\Sigma_\pm\equiv {\rm tr}\,(h_\pm\, {}^*\Sigma)/2$,
 and therefore the damping rate (\ref{gpm}).

Note that the simplified vertex (\ref{Gproj}) has no discontinuity,
so that the whole imaginary part of $ {}^*\Sigma$ in the
kinematical regime of interest arises by cutting the internal
lines in Fig.~\ref{figeff}.a. Specifically, the previous approximations
yield the dominant (infrared singular) piece of the one-loop damping rate 
 as\cite{Pisarski93,Rebhan95}
\beq\label{gpm*}
\gamma_\pm(p) 
 &\simeq&  z_\pm\,g^2  T \int \frac{{\rm d}^3q}{(2\pi)^3}\,
\frac{v^i_\pm}{z_\pm}
\,\frac{\delta^{ij}}{q^2}\,\frac{v^j_\pm}{z_\pm}
\, {\rm Im}\,\,\frac
{-z_\pm}{\omega -\omega_\pm(p) - |v_\pm|
q\cos\theta +i\eta}\nonumber\\& \simeq &
\alpha T|v_\pm(p)|
\ln \frac{\omega_p}{|\omega-\omega_\pm(p)|},
\eeq
which is very close to eq.~(\ref{gammaIR}) for a hard fermion
(recall that $|{\bf v}| = 1$ for the hard quasiparticle).

Consider now the higher order corrections to $\gamma$,
with emphasis on the leading infrared contributions.
By relying mostly on the gauge symmetry, we argue now
that the most singular contributions
to $\gamma$ arise from multi-loop diagrams which involve
the (resummed) 3-point photon-fermion vertex, {\it but not the
higher order vertices}\footnote{This can be also verified
by power counting, as in Ref. \cite{Pisarski93}
for the one-loop calculation. Namely, cutting a vertex rather than
a fermion propagator, yields a factor 
of $1/({\bf v}_{s}(p)\cdot {\bf q})$ less, and thus a
less singular infrared behavior.}.
This is so since
in the kinematical regime of interest, the inverse fermion propagator,
\beq {}^*\Delta_s^{-1}(\omega,{\bf p+q})\simeq
-\Bigl(\omega-\omega_{s}(p)
-{\bf v}_{s}(p)\cdot {\bf q}\Bigr)\,\frac{1}{z_{s}(p)}\,, \eeq
 is {\it linear} in
the photon momentum ${\bf q}$, so that the
Ward identity (\ref{wid}) can be satisfied by a 3-point vertex
 ${}^*\Gamma^i({\bf p,p+q})$ which is {\it independent}
of the  momentum of the photon leg. And we have seen indeed
that the singular one-loop contribution is obtained
by replacing  ${}^*\Gamma^i({\bf p,p+q})$ with
 ${}^*\Gamma^i({\bf p,p})$, which is independent of ${\bf q}$
and (up to a spin projector) equal to $v^i_s(p)/z_s(p)$.
Furthermore, with a ${\bf q}$-independent 3-point vertex,
all the other, higher, Ward identities --- as the one shown
in eq.~(\ref{wid2}) ---
are trivially satisfied by setting the $n$-point HTL's
with $n\ge 4$ to zero. Since, as alluded to before,
the vertex HTL's are essentially determined by the
Ward identities, it follows that the higher-point
vertices (beyond the 3-point function) are not
important in the kinematical regime of interest.

We thus conclude that, in order to isolate the most singular contributions to
$\gamma_s$ ($s=\pm$) in perturbation theory, we have to consider the
{\it same} diagrams as for
the hard fermion, and evaluate them with
 the following simplified Feynman rules:\\
(i) the photon propagator   $D_0^{ij}({\bf q})=
\delta^{ij}/q^2$;\\
(ii) the fermion propagator ${}^*\Delta_{s}(\omega,{\bf p+q})$
from eq.~(\ref{res});\\
(iii) the photon-fermion vertex 
${}^*\Gamma^i_s (p)= v^i_s(p)/z_s(p)$.\\
The momentum integrals over the photon momenta
 should be computed with an upper
 cut-off of the order $\omega_p$.  Strictly speaking,
the above simplifications hold only for very soft 
momenta, $q\ll p\sim gT$, and not up to momenta
 $q\sim \omega_p$. This is not important, however,
since the dominant (singular) contributions arise
from the limit $q\to 0$ and are insensitive to the upper
cut-off. 

Note that the above Feynman rules are essentially those
of a {\it local} effective field theory, in contrast
with the general HTL Feynman rules, which are non-local.
(The apparent dependence on $\omega$ and $p$ is irrelevant
here, since these are the {\it fixed} energy and momentum of
the external line; they enter the computation as parameters).
 Furthermore,  the IR contribution to $\gamma$
is largerly insensitive to the details of
the HTL resummation, which enters only via the global
factors $ v_s(p)$ and $z_s(p)$. Actually, to the order
of interest, $\gamma_s$ is even independent of the residue
$z_s(p)$, as also suggested by the one-loop result (\ref{gpm*}).
This is so since a general $n$-loop graph contributing
to $\Sigma_s$ (in the simplified perturbation theory
introduced above) involves $2n$ vertices 
${}^*\Gamma^i_s$, and therefore a factor $z_s^{-2n}$,
and $(2n-1)$ propagators ${}^*\Delta_s$, which yield a factor
$z_s^{2n-1}$. The remaining factor of $1/z_s$ disappears
in the computation of $\gamma_s = -z_s {\rm Im}\,\Sigma_s$.
 
\subsection{The Bloch-Nordsieck model for a soft fermion}

At this point, the analysis of the
dominant mass-shell behavior of the soft fermion becomes
almost identical to the corresponding analysis for the hard fermion.
This this analogy is due to the fact that
the soft photons responsible for the IR divergences
have typical momenta $q\ll gT$, which are much smaller than
the momentum $p \sim gT$ of the soft fermion. In view of this, the
 whole discussion in sections 3 and 4 can be directly extended
 to the case of a soft fermion. 

Specifically, the simplified Feynman rules which apply
in the IR regime are, once again, those of the {\BN} model,
and can be summarized in the following functional integral
representation of the soft fermion propagator:
\beq\label{SS}
S_\pm(x,y)= Z_0^{-1}\int [{\rm d} {\bf A}]\, G_\pm(x,y|{\bf A})\,{\rm
exp}\left\{-\frac{1}{2}\,
\Bigl({\bf A}, D_0^{-1}{\bf A}\Bigr)_0\right\}.\eeq
In this equation,  $ G_s(x,y|{\bf A})$
 is the {\BN} propagator for the quasiparticle in the mode
 $s$,  $s=\pm$, in the presence of the
static magnetic field $A^\mu=(0,{\bf A}({\bf x}))$, and satisfies
(with $v_s^\mu\equiv (1, {\bf v}_s)$)
\beq\label{GAS}
-i\,(v_s \cdot D_x)\,G_s(x,y|A)&=&z_s\,\delta^{(4)}(x-y).\eeq
Furthermore,
\beq\label{SEFFS} 
\Bigl({\bf A},D_0^{-1}{\bf A}\Bigr)_0 \,=\,\frac{1}{T} \int {\rm
d}^3x\, {\rm d}^3y\,A^i({\bf x})\,
D^{-1}_{0\,ij} ({\bf x}-{\bf y}) A^j ({\bf y}),\eeq
where the vector field $A^i({\bf x})$ has been defined
in eq.~(\ref{Ast}), and $D_{0\,ij}({\bf x})=\delta_{ij}/4\pi x$.
Note that the free (retarded) BN propagator, as obtained from
eq.~(\ref{GAS}) with $A=0$, reads
\beq\label{G0S}
G_s(\omega, {\bf p+q})\,=\,\frac{-z_s}
{\omega-{\bf v_s \cdot (p+q)} +i\eta}\,.\eeq
Strictly speaking,  the mass-shell
for the BN particle of momentum $p$, that is
 $\omega={\bf v_s \cdot p}$,
is different from the real leading-order mass-shell, at
$\omega=\omega_s(p)$. This is so, of course, since
the dispersion relations for soft fermions are not linear,
so that the group velocity $|{\bf v_s}|$ is really momentum dependent.
However, this difference is not important,
since the BN propagator (\ref{G0S})
presents the correct dependence on ${\bf q}$ in the mass-shell
limit. Compare in this respect eqs.~(\ref{G0S}) and (\ref{res}):
in both these equations, it is the difference
in energy with respect to the mass-shell which matters,
rather than the precise value of the mass-shell energy itself.
For $\omega=\omega_s(p)$ in eq.~(\ref{res}), and respectively
for  $\omega={\bf v_s \cdot p}$ in eq.~(\ref{G0S}),
the propagators in these two equations become identical.

 Eqs.~(\ref{SS})--(\ref{GAS}) are further
manipulated as in section 3.5 (recall, especially, the discussion
after eq.~(\ref{FS})). As a result, we obtain
 the retarded propagator for the two fermionic modes 
$\pm$, for momenta $p\sim gT$ and energies close to the mass-shell,
$\omega\simeq \omega_{\pm}(p)$. It reads
\beq\label{SRpm}
S_{\pm}(\omega, {\bf p}) &=&i\,z_{\pm}(p)\int_0^{\infty}{\rm d}t
\,{\rm e}^{it(\omega- \omega_{\pm}(p)+i\eta)}\,\Delta_{\pm}(t),
\nonumber\\ \Delta_{\pm}(t)&= &
 {\rm exp}\left \{-g^2T
\int \frac{{\rm d}^3q}{(2\pi)^3} 
\frac{\tilde D_\pm({\bf q})}{({\bf v_\pm\cdot q})^2}\,
\Bigl(1-  {\rm cos}\,t({\bf v_\pm\cdot q})\Bigr)\right\}.\eeq
In this equation, $\tilde D_\pm({\bf q})=v^i_\pm D^{ij}_0({\bf q})
v^j_\pm = v^2_\pm/q^2$, so that we can write
\beq\label{DD}&
\Delta_{\pm}(t)= &\Delta (|v_{\pm}|t),\eeq
with $\Delta(t)$ as given by eq.~(\ref{SR0})
where $\tilde D({\bf q}) \to 1/q^2$ and ${\bf v}$ is an arbitrary 
unit vector. Note that the functions $\Delta_{\pm}(t)$ are 
implicitly  dependent on the momentum $p$, via the 
group velocities $v_\pm(p)$.
 Both  the mass-shell behavior of the propagator (\ref{SRpm})
and the large time behavior of the
propagator $S_\pm(t,{\bf p})$ follows from the  analysis
in section 4.  At very large times $\omega_p |v_\pm|t
\gg 1$, we have
\beq\label{DLTS}
\Delta_\pm(\omega_p |v_\pm|t\gg 1)\simeq
 {\rm exp}\Bigl\{ -\alpha T|v_\pm|t \ln (\omega_p
|v_\pm| t)\Bigr\}.\eeq
The spectral density of the mode $s$ is peaked around $\omega=\omega_s(p)$,
with a width of the order $g^2T|v_s| \ln (1/g)$.
In particular, for the lower mode $\omega_-(p)$, and for
$p=p_c$, where $v_-(p_c)=0$, eq.~(\ref{DD}) shows that,
to this approximation, the ``plasmino'' mode is not damped, in accordance
with the one-loop result for the damping rate,
eq.~(\ref{gpm*}).

\section{Conclusions}

The analysis presented in this paper suggests that the damping of
the fermionic excitations with momenta $p\gg g^2T$
is {\it not} exponential in time, but of the more
complicated form $S_R(t)\sim
{\rm e}^{-iE(p)t}\exp\{-\alpha T |v|\, t\, \ln(\omega_p|v|t)\}$, 
where ${\bf v}=\del E/ \del {\bf p}$ is the group velocity of
the excitation,  $\omega_p \sim gT$ is the plasma frequency,
and $\alpha=g^2/4\pi$. As a consequence, the 
 retarded propagator $S_R(\omega)$ has no
 quasiparticle pole,  but the spectral density shows
nevertheless a sharp
resonance peaked at $\omega=E(p)$,  with a width $\sim g^2T\ln(1/g)$.
At the present level of accuracy, the mean energy $E(p)$ is given
by the leading-order approximation, namely $E(p)=p$ for a hard
excitation, and $E(p)=\omega_\pm(p)$ for a soft one.
We note that this result  solves the IR problem of the
damping rate in a very ``soft'' way,
by essentially replacing the IR cut-off $\mu$
in the perturbative result $\Delta_L(t)
=\exp\{-\alpha T |v|\, t\, \ln(\omega_p|v|/\mu)\}$
with the inverse of the time. Thus, quantitatively, the lifetime
of the excitation does not differ much from that obtained from
leading-order perturbation theory.

The asymptotic behaviour of the retarded propagator
 has been obtained by solving exactly an effective
theory which reproduces all the leading infrared divergences of the
perturbation theory. The physical processes which  are 
responsible for these divergences are the
 multiple  collisions  involving the exchange of
long wavelength, quasi-static, magnetic photons, which
are not screened by plasma effects. 
By comparison, the longitudinal, gauge sector is less singular
in perturbation theory, and does not contribute 
to the dominant large time behaviour of the non-perturbative solution.

At finite temperature,
the presence of the thermal bath amplifies the IR divergences,
in such a way that they become effectively those
of a three-dimensional gauge theory.
Then, a comparison with massive ${\rm QED}_3$ \cite{Sen}
helps explaining why an IR divergence occurs
for the one-loop damping rate, in
contrast to the zero temperature case where
the IR problem does not affect  the dispersion equation,
but only the residue of the propagator\cite{Bogoliubov}.
At this point, we should recall that
 the explicit solution that we have proposed here
relies essentially on the 3-dimensional character of the dominant singularities. 
This has been widely recognized in relation with
the infrared structure of thermal field theories \cite{Linde80},
and, in the calculation of
{\it static} quantities (like the free energy or the screening masses),
it has been exploited in the method of ``dimensional reduction''
(see \cite{Nadkarni83,Braaten94,Kajantie94,debye} and references therein).  
We emphasize, however, that the damping rate is a dynamical quantity,
and the usefulness of the dimensional reduction for this problem
is not a priori obvious, given the subtleties of the analytic
 continuation from Matsubara to real external energy.
If a dimensional reduction occurs 
in the computation of the large time behaviour, this is because of
 the particular IR behaviour of the magnetic photon propagator,
 as displayed in eqs.~(\ref{singDT}) or (\ref{rhot}).
The dynamical information which is contained in the later equations
refers not only to the absence of the magnetic screening,
but also to the phenomenon of Landau damping.

It is also worth emphasizing that  our result takes into account only
 the most singular terms of the perturbative expansion.
Because of the approximation used, we have lost control on the subleading terms.
Although,  in a strict perturbative sense,
 these are a priori less important,
one cannot completely exclude the possibility that they may
 still modify our results in a qualitative way.
It is hard to see however how they could destroy the quasiparticle
picture, which we have shown to survive after a complete
treatement of the leading IR divergences.
Improvements of our solution may require an appropriate generalisation
of the Bloch-Nordsieck model to finite temperature,
a task that we have explored in this paper, but
 without reaching a definite conclusion.
There are at least two points where the 
thermal BN model could possibly complete our previous
analysis:  the dynamical emergence of the upper cut-off
$\sim gT$ (recall that, in the effective 3-dimensional theory,
this cut-off has been introduced by hand),
and, related to this, the consistent computation of the subleading terms 
beyond $\ln(\omega_p t)$ in eq.~(\ref{DLT}),
 that is, the terms of order $g^2 T$
 which multiply the time in the exponent of $\Delta(t)$.

It is finally natural to ask what is the relevance
of the present solution for the non-Abelian QCD plasma.
It is widely believed that the self-interactions
of the chromomagnetic gluons 
may generate  magnetic screening at the scale $g^2 T$.
As a crude model, we may include a screening mass $\mu\sim g^2T$
in the magnetostatic propagator in the QED calculation.
 Then eq.~(\ref{SR0}) provides,
 at very large times $t\simge 1/g^2T$,
an exponential decay,  $\Delta_\mu(t)
\sim \exp(-\gamma t)$ with $\gamma = \alpha T\ln(\omega_p/\mu)
=  \alpha T\ln(1/g)$.
However, in the physically more interesting regime of intermediate
times $1/gT \ll t \ll 1/g^2 T$, the behavior is governed
uniquely by the plasma frequency, according to our result
(\ref{DLT}): $\Delta_\mu(t)\sim \exp ( -\alpha Tt \ln \omega_p
t)$.  Thus, at least within this limited model,
which is QED with a ``magnetic mass'', the time behavior
in the physical regime remains controlled by the
Bloch-Nordsieck mechanism. But, of course, this result gives no
serious indication about the real situation in QCD, since
it is unknown whether, in the present problem,
 the effects of the gluon self-interactions
can be simply summarized in terms of a magnetic mass.

\vspace*{1.5cm}
{\noindent {\large{\bf Acknowledgements}}}

During the elaboration of this paper, we have benefited
from discussions and useful remarks from a number
of people. It is a pleasure to thank
R. Baier, G. Baym, M. LeBellac, B. M\"uller, R.D. Pisarski,
A.K. Rebhan, D. Schiff and B. Vanderheyden.

\setcounter{equation}{0}
\vspace*{2cm}
\renewcommand{\theequation}{A.\arabic{equation}}
\appendix{\noindent {\large{\bf Appendix A}}}

In this Appendix, we collect the sum rules for the
photon spectral densities which are used in section 2.2.

The electric and magnetic spectral densities are
defined  in eq.~(\ref{rhos}) in terms of the corresponding
propagators. In the hard thermal loop approximation,
they involve both pole and cut pieces, as shown in eq.~(\ref{RRHO}).
They satisfy the following sum-rules\cite{Pisarski93},
which trade the integrals over the off-shell spectral
densities $\beta_{l,\,t}(q_0,q)$
 for functions of $\omega_s(q)$ and $z_s(q)\,:$
\beq\label{SumR}
\int_{-q}^q\frac{{\rm d}q_0}{2\pi q_0}\,
\beta_l(q_0,q)&=&\frac{1}{q^2}\,-\,\frac{1}{q^2+m_D^2}\,-\,
\frac{z_l(q)}{\omega_l^2(q)},\nonumber\\
\int_{-q}^q\frac{{\rm d}q_0}{2\pi q_0}\,
\beta_t(q_0,q)&=&\frac{1}{q^2}\,-\,\frac{z_t(q)}{\omega_t^2(q)},
\nonumber\\
\int_{-q}^q\frac{{\rm d}q_0}{2\pi}\, q_0\,
\beta_t(q_0,q)&=&1\,-\,{z_t(q)}.\eeq
The first two of these sum rules are obtained by simply setting $\omega=0$
in the spectral representations (\ref{Sspec}), and by using
${}^*\Delta_l(0,q)=-1/(q^2+m_D^2)$,
${}^*\Delta_t(0,q)=-1/q^2$, together with eq.~(\ref{RRHO}).
As for the third one, this is obtained
by inserting eq.~(\ref{RRHO}) into the familiar sum-rule
\beq
\int_{-\infty}^\infty\frac{{\rm d}q_0}{2\pi}\, q_0\,
{}^*\rho_t(q_0,q)&=&1,\eeq
which is a consequence of the equal-time commutation relation
for the quantum fields\cite{MLB96}.

The use of the sum rules (\ref{SumR}) is convenient to study
both the ultraviolet and the infrared behavior of the $q$-integral 
in eq.~(\ref{g5}). To this aim, we need the dispersion relations
$\omega_{l,\,t}(q)$ \cite{Klimov81,Weldon82,Pisarski89a}
 and the corresponding residues
$z_{l,\,t}(q)$, which, in our conventions, read:
\beq\label{resexp}
z_t\,=\,\frac{2\omega_t^2\,(\omega_t^2-q^2)}
{3\omega_p^2 \omega_t^2-(\omega_t^2-q^2)^2}\,,\qquad\qquad
z_l\,=\,\frac{2\omega_l^2 (\omega_l^2/q^2-1)}
{3\omega_p^2-(\omega_l^2-q^2)}\,.\eeq 
At large momenta, $q\gg \omega_p$, we
have the approximate expressions\cite{Pisarski89a}
\beq\label{largeq}
\omega^2_t(q)\simeq q^2 +3\omega_p^2/2,\qquad &\,& \qquad
\omega^2_l(q)\simeq q^2(1+4x_l(q)),\nonumber\\
z_t(q)\simeq 1-\,  \frac{3\omega_p^2}{4q^2}\left(
\ln  \frac{8q^2}{3\omega_p^2} -3\right),\qquad &\,& \qquad
z_l(q)\simeq \frac{8q^2}{3\omega_p^2}\,x_l(q)\,,\eeq
where
\beq x_l(q)\equiv \exp\left( -\frac{2q^2}{3\omega_p^2}\,-2\right).\eeq
From eqs.~(\ref{SumR}) and (\ref{largeq}), we obtain,
for $q\gg \omega_p$ (recall that $m_D^2=3\omega^2_p$),
\beq\label{SumRUV}
\int_{-q}^q\frac{{\rm d}q_0}{2\pi q_0}\,
\beta_l(q_0,q)&\simeq &\frac{3\omega_p^2}{q^4}\,,\nonumber\\
 \int_{-q}^q\frac{{\rm d}q_0}{2\pi q_0}
\left(1-\frac{q_0^2}{q^2}\right)
\beta_t(q_0,q)&\simeq &\frac{3\omega_p^2}{2 q^4}\,.\eeq
These estimates show that  the integrand in eq.~(\ref{g5}) behaves like
$\omega_p^2/q^3$ for momenta  $q\gg \omega_p$.

We turn now to  momenta $q\ll \omega_p$. We then have
\beq\omega^2_t(q)\simeq \omega_p^2 + 6q^2/5,\qquad &\,& \qquad 
\omega^2_l(q)\simeq \omega_p^2 + 3q^2/5,\nonumber\\
z_t(q)\simeq 1-\frac{q^2}{5\omega_p^2}\,,\qquad &\,& \qquad
z_l(q)\simeq \frac{\omega_p^2}{q^2}\left(1+
{\cal O}(q^4/\omega_p^4)\right),\eeq
so that 
\beq\label{SumRIR}
\int_{-q}^q\frac{{\rm d}q_0}{2\pi q_0}\,
\beta_l(q_0,q)&\simeq &\frac{4}{15}\,
\frac{1}{\omega_p^2},\nonumber\\
\int_{-q}^q\frac{{\rm d}q_0}{2\pi q_0}\,
\beta_t(q_0,q)& \simeq &\frac{1}{q^2}\,-\,
\frac{1}{\omega_p^2},\nonumber\\
\int_{-q}^q\frac{{\rm d}q_0}{2\pi}\, q_0\,
\beta_t(q_0,q)&\simeq &\frac{1}{5\omega_p^2}\,.\eeq
When these expressions are inserted in eq.~(\ref{g5}),
 the contribution in $1/q^2$ of the magnetic
spectral function (the second line in eq.~(\ref{SumRIR}))
 generates a logarithmic IR singularity.

\setcounter{equation}{0}
\vspace*{2cm}
\renewcommand{\theequation}{B.\arabic{equation}}
\appendix{\noindent {\large{\bf Appendix B}}}

Since there is no phase-space available for the direct decay of
the on-shell fermion into a pair of massless particles, one expects that the
damping rate computed from the
{\it bare} one-loop fermion self-energy should vanish.
However, at finite temperature, this  argument
 is complicated by infrared singularities which arise because
of the enhancement of collinear singularities
by the Bose-Einstein thermal factor. 

To illustrate this problem, we consider the calculation of the damping rate
to bare one-loop order in the Coulomb gauge. This is obtained 
by simply replacing, in eq.~(\ref{g1}), the photon spectral functions
with their bare counterparts, namely $\rho_l^{(0)}=0$
and $\rho_t^{(0)}(q_0,q) = \rho_0(q_0,q)$, with $\rho_0$
from eq.~(\ref{rho0}). In the on-shell limit, the whole
 contribution to $\gamma$ comes from space-like photons, with
$|q_0|\le q$. However, since the free spectral density (\ref{rho0})
has support precisely at the  integration limits $q_0=\pm q$,
we should be more careful when evaluating eq.~(\ref{g1})
in the on-shell limit $\omega\to p$.
 For $\omega$ close to, but different from, $p$,
the latter equation yields (compare with eq.~(\ref{g4}))
\beq\label{g2}
\gamma_0(\omega\simeq p)\simeq \pi g^2 T\int \frac{{\rm d}^3q}{(2\pi)^3}
 \int_{-\infty}^\infty \frac{{\rm d}q_0}{2\pi q_0}\,
\, \delta(\omega-p -q_0 +q\cos \theta)
(1-\cos^2\theta)\rho_0(q_0,q).\eeq
After the angular integration, we obtain
\beq\label{g3}
\gamma_0(\omega\simeq p)&\simeq& \frac{g^2 T}{8\pi}
(\omega-p) \int_0^\infty \frac{{\rm d}q}
{q^2} \int_{\omega-p-q}^{\omega-p+q}\frac{{\rm d}q_0}{ q_0}\,
(2q_0- (\omega-p))\Bigl[\delta(q_0-q)-
\delta(q_0+q)\Bigr]\nonumber\\
&=& \frac{g^2 T}{4\pi}
\,|\omega-p| \int_{|\omega-p|/2}^\infty \frac{{\rm d}q}
{q^2}\left(1-\frac{|\omega-p|}{2q}\right).\eeq
If we let now $\omega\to p$, the factor in the front of the last
integral goes to zero, but the integral itself becomes IR divergent.
An explicit calculation shows that 
the r.h.s. of eq.~(\ref{g3}) is in fact independent of $(\omega-p)$,
and equals
\beq\label{g31}
\gamma_0 \,=\, \frac{g^2 T}{4\pi}.\eeq
This result is, however, unphysical. It arises from the
emission or the absorbtion of {\it collinear}
($\theta = 0$, $q_0=q$ or, respectively,
 $\theta=\pi$, $q_0=-q$) massless photons, whose
contributions are enhanced by the Bose-Einstein factor
$T/q_0$. Such contributions do not survive screening corrections.
However, since  the gauge-dependent terms in the photon propagator
are not modified by the plasma effects, they may generate
 --- by the mechanism alluded to before --- a non-vanishing
contribution to the on-shell self-energy \cite{Schiff92}.
 Note that an entirely similar
problem arises in the three-dimensional gauge theories
at zero-temperature, when computing the dispersion equation
to one loop order\cite{Sen}. 

To overcome this problem, it has been suggested \cite{Sen,Rebhan93}
to take the  on-shell limit in the presence of an IR regulator,
say, an IR cut-off $\mu$. With such a cut-off, the $q$-integral 
in eq.~(\ref{g3}) remains finite as $\omega\to p$, and
the total result for $\gamma_0(\omega =p)$ vanishes.
Thus, the damping rate remains zero at the bare one-loop level,
 as expected. In the same way one verifies that the 
dispersion relation is gauge-independent, as it should\cite{KKR90}.
On the other hand, the residue of the propagator at the mass-shell
becomes dependent on the IR cut-off $\mu$, and (linearly) divergent
as $\mu\to 0$.

\setcounter{equation}{0}
\vspace*{2cm}
\renewcommand{\theequation}{C.\arabic{equation}}
\appendix{\noindent {\large{\bf Appendix C}}}

We verify here, on an explicit two-loop calculation, some 
general features of the infrared behavior of the on-shell
self-energy in perturbation theory. Specifically,
we shall show that the leading divergences are power-like,
and can be fully taken into account by restricting all the internal
Matsubara sums to their zero frequency photon modes.

\begin{figure}
\protect \epsfxsize=14.cm{\centerline{\epsfbox{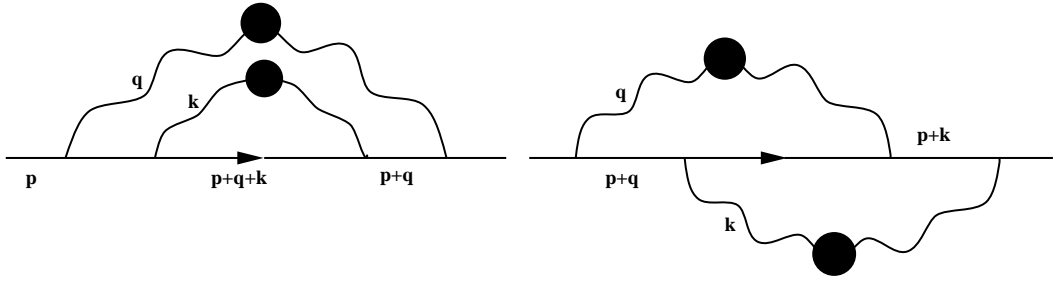}}}
	 \caption{Two-loop diagrams for the fermion self-energy.}
\label{eff2}
\end{figure}
At two loop order, the fermion self-energy is given by
the two diagrams in Fig.~\ref{eff2},  which yield
\beq\label{Sig2L}
\Sigma^{(2)}(p)&=&-\,(g^2 T)^2\sum_{q^0= i\omega_m}\sum_{k^0= i\omega_r}
 \int \frac{{\rm d}^3q}{(2\pi)^3}  \int \frac{{\rm d}^3k}{(2\pi)^3}\,
\,\gamma_\mu S_0(p+q)\gamma_\rho S_0(p+q+k)
\nonumber\\ &\,& \qquad\qquad \Bigl[
\gamma_\lambda S_0(p+q) \gamma_\nu + \gamma_\nu  S_0(p+k)
 \gamma_\lambda\Bigr] \,{}^*D^{\mu\nu}(q)\,
{}^*D^{\rho\lambda}(k)\,.\eeq
 The notations here are as in eq.~(\ref{Sigeff}); for instance,
$p^0= i\omega_n= i(2n+1)\pi T$,  $\,q^0= i\omega_m= i2\pi mT$ 
and  $k^0= i\omega_r= i2\pi rT$, with integers $n$, $m$ and $r$.
According to eq.~(\ref{PIpm}), the correction to the positive mass-shell 
is determined by the function
$\Sigma_+(\omega,p) = {\rm tr}\,\Bigl(
h_\pm(\hat {\bf p}) \Sigma(\omega, {\bf p})\Bigr)/2$.
As $\omega \simeq p$, the most singular contributions to $\Sigma_+$
are obtained by using the effective Feynman rules 
described at the end of section 3.1. At two loop level, this amounts
to replacing eq.~(\ref{Sig2L}) by
\beq\label{PI2L}
\Sigma_+^{(2)}(p)&=&-\,(g^2 T)^2\sum_{q^0= i\omega_m}\sum_{k^0= i\omega_r}
 \int \frac{{\rm d}^3q}{(2\pi)^3}  \int \frac{{\rm d}^3k}{(2\pi)^3}\,
\, G_0(p+q)\,G_0(p+q+k)
\nonumber\\ &\,& \qquad\qquad\qquad \Bigl[
G_0(p+q) + G_0(p+k)\Bigr] \tilde D(q)\,\tilde D(k)\,,\eeq
where $\tilde D(q) \equiv v^i\,{}^*D^{ij}(q) v^j$
 is the (HTL resummed)  propagator 
of the magnetic photon. (We recall that the electric
propagator does not yield IR singularities.)
Eq.~(\ref{PI2L}) is precisely the two-loop
self-energy in the Bloch-Nordsieck approximation.

The Matsubara sums over $\omega_m$ and $\omega_r$
are conveniently performed by contour methods, and by using
the spectral representation (\ref{Sspec}) of ${}^*\Delta_t(q)$.
In doing this, one gets several terms, corresponding to the poles
of the various propagators in the complex planes $q^0$ and $k^0$.
Every such a term involves three energy denominators,
and the product of two statistical factors. The latter
are either of the bosonic or of the fermionic type,
according to whether they correspond to poles of
a photon propagator, or of an electron propagator, respectively.
When the external energy approaches the tree-level mass-shell,
$\omega \to p\equiv {\bf v\cdot p}$, all the energy denominators
are soft, of the type $1/(q_0-{\bf v \cdot q})$, and may give
 infrared problems. (The hard energy denominators,
which were  potentially present in the full two-loop self-energy
(\ref{Sig2L}), have been eliminated by the simplified
Feynman rules leading to eq.~(\ref{PI2L}).)
Then, the {\it leading} IR singularities arise uniquely
from the terms which involve the product of {\it two Bose-Einstein
distribution functions}, since
$N(q^0) N(k^0) \simeq T^2/(q^0 k^0)$ at soft momenta.
By isolating these most singular terms, we obtain, after
a straightforward calculation,
\beq\label{PIsing}
\Sigma_+^{(2)}(\omega\simeq p)&\simeq& (g^2 T)^2
 \int \frac{{\rm d}^3q}{(2\pi)^3}  \int \frac{{\rm d}^3k}{(2\pi)^3}\,
\int_{-\infty}^\infty\frac{{\rm d}q_0}{2\pi q_0}\,
\,{}^*\rho_t(q_0,q)\,\int_{-\infty}^\infty\frac{{\rm d}k_0}{2\pi k_0}\,
\,{}^*\rho_t(k_0,k)\nonumber\\ &\,&\qquad
\frac{1}{\omega + q_0 -{\bf v}\cdot ({\bf p+q})}\,
\frac{1}{\omega + q_0 +k_0 -{\bf v}\cdot ({\bf p+q+k})}\nonumber\\ &\,&\qquad
\left[\frac{1}{\omega + q_0 -{\bf v}\cdot ({\bf p+q})}\,+\,
\frac{1}{\omega + k_0 -{\bf v}\cdot ({\bf p+k})}\right]
\,,\eeq
where it is understood that the external energy carries
a small positive imaginary part ($\omega \to \omega + i\eta$).

The energy integrals over $q_0$ and over $k_0$ involve
both the pole and the cut pieces of the photon spectral
density. However, it is only the off-shell (or cut) piece
of ${}^*\rho_t$ which yields a singular contribution,
so we may as well restrict the aforementioned energy integrals
to space-like momenta, $|q_0|\le q$ and
$|k_0|\le q$, and replace the full spectral functions
by $\beta_t$. Then, the subsequent analysis follows closely the discusion of the
(resummed) one-loop self-energy in section 2.2.
The singular domain is that of very soft photon momenta,
$q,\,k\ll gT$, where we can use eq.~(\ref{rhot}) to replace
$\beta_t(q_0\ll q)/q_0$ by $(2\pi/q^2)\delta(q_0)$.
At the same time, we have to supplement the momentum integrations
with an upper cut-off of the order of $\omega_p\sim gT$.
The net effect is that the leading singular piece
of $\Sigma_+^{(2)}(\omega\simeq p)$ is the same as it would be obtained
by retaining only the static terms 
 $\omega_m=\omega_r=0$ in the Matsubara sums of eq.~(\ref{PI2L}).
That is,
\beq\label{PIS}
\Sigma_+^{(2)}(\omega\simeq p)&\simeq& (g^2 T)^2
 \int \frac{{\rm d}^3q}{(2\pi)^3}\,\frac{1}{q^2}\,  \int \frac{{\rm d}^3k}{(2\pi)^3}\,
\frac{1}{k^2}\,
\frac{1}{\omega  -{\bf v}\cdot ({\bf p+q+k})}\,\nonumber\\ &\,&\qquad
\frac{1}{\omega -{\bf v}\cdot ({\bf p+q})}
\left[\frac{1}{\omega  -{\bf v}\cdot ({\bf p+q})}\,+\,
\frac{1}{\omega  -{\bf v}\cdot ({\bf p+k})}\right]
\,.\eeq
Since this is divergent as $\omega \to {\bf v\cdot p}$, we
take the mass-shell limit in the presence on an IR cut-off $\mu$,
and obtain
\beq\label{PIS1}
\Sigma_+^{(2)}(\omega\simeq p)\simeq - (g^2 T)^2
 \int \frac{{\rm d}^3q}{(2\pi)^3}\,\frac{1}{q^2}\,
\frac{1}{({\bf v\cdot q}-i\eta)^2}\,
  \int \frac{{\rm d}^3k}{(2\pi)^3}\,\frac{1}{k^2}\,
\frac{1}{{\bf v\cdot k} - i\eta}\,
=\,i \,\frac{2}{\pi}\,\frac{(\alpha T)^2}{\mu}\,\ln
\frac{\omega_p}{\mu}\,.\nonumber\\ \eeq
We thus find the linear plus logarithmic infrared divergence
mentioned in section 2.3. 

According to eq.~(\ref{domeg}), the computation of $\gamma^{(2)}$ ---
 the two-loop contribution to the damping rate --- requires also
 the one-loop residue,
$ z^{(1)}(p) - 1 =(\del \Sigma_+^{(1)}/\del\omega)\,$.
Similarly to eq.~(\ref{PIS}), we obtain the
leading IR-singular contribution to $\Sigma_+^{(1)}$ in the form
\beq\label{PI1L}
\Sigma_+^{(1)}(\omega, p)&\simeq& g^2 T
 \int \frac{{\rm d}^3q}{(2\pi)^3}\,\frac{1}{q^2}\, 
\frac{1}{\omega -{\bf v}\cdot ({\bf p+q}) +i\eta}\,,\eeq
and thus
\beq\label{zz}  z^{(1)}(p) - 1 \simeq - g^2 T
  \int \frac{{\rm d}^3q}{(2\pi)^3}\,\frac{1}{q^2}\,
\frac{1}{({\bf v\cdot q}-i\eta)^2}\,
\simeq \,\frac{2}{\pi}\,\frac{\alpha T}{\mu}\,, \eeq
in the presence of the IR regulator. The linear IR
divergence of the residue compensates the dominant
singularity of the two-loop self-energy (\ref{PIS1}),
so that the leading contribution 
to  $\gamma^{(2)}$ --- which remains beyond the
accuracy of the present computation --- is of the
order $(\alpha^2 T^2/\omega_p) (\ln
(\omega_p/\mu))^2\sim g^3 T  (\ln
(\omega_p/\mu))^2$. Even if still divergent as $\mu \to 0$,
this does not contribute to the order $g^2T$
 which is our concern here.

Let us finally provide an all order argument for
the cancellation of the strongest, power-like, infrared divergences
in the perturbative evaluation of $\gamma$.
To this aim, we consider the Dyson-Schwinger
equation for the fermion self-energy within
the effective three-dimensional Bloch-Nordsieck theory:
\beq\label{DSBN}
 \Sigma(\omega,{\bf p})\,=\,-g^2T \int \frac{{\rm d}^3q}{(2\pi)^3} 
\, v^i\,S(\omega,{\bf p+q})\,\Gamma^j({\bf p+q, p}) 
\,D^{ij}_0({\bf q}),\,\eeq
where $D^{ij}_0({\bf q})= \delta^{ij}/q^2$,
$S$ is the full BN propagator,
\beq \label{SBN}
S(\omega,{\bf p+q})\,=\,
\frac{-1}{\omega - {\bf v}\cdot ({\bf p+q})
- \Sigma(\omega,{\bf p+q})}\,,\eeq
and $\Gamma^j({\bf p+q, p}) $ is the full vertex,
which is related to $S$ via the Ward identity
\beq\label{wid3}
q^j\,\Gamma_j({\bf p+q,p})= S^{-1}(\omega,{\bf p+q})
-S^{-1}(\omega,{\bf p})\,.\eeq
We make now the usual assumption \cite{BallChiu}
 that the dominant IR behavior
involves only the longitudinal piece of the vertex. This is
 entirely determined by the Ward identity:
\beq\label{long}
\Gamma^j({\bf p+q, p})\,=\,\frac{v^j}{{\bf v\cdot q}}\left
[S^{-1}(\omega,{\bf p+q})
-S^{-1}(\omega,{\bf p})\right].\eeq
 When inserted in eq.~(\ref{DSBN}), this yields
\beq\label{DS2}
 \Sigma(\omega,{\bf p})\,\simeq\,-g^2T \int 
\frac{{\rm d}^3q}{(2\pi)^3} \,\frac{1}{q^2}\,
\frac{1}{{\bf v\cdot q} - i\eta}\Bigl[1\,-\,
S^{-1}(\omega,{\bf p})\,S(\omega,{\bf p+q})\Bigr]\,.\eeq
As already explained, eq.~(\ref{DSBN}) reproduces
the most singular terms of the perturbative expansion,
and this remains true after inserting the approximation
(\ref{long}) for the vertex function,
as can be verified explicitly by developing
eq.~(\ref{DS2}) in perturbation theory.
We now take the on-shell limit in the presence
of an IR cut-off $\mu$, taken as a small
photon mass. As long as $\mu\ne 0$, there is no IR
problem, and we expect the mass-shell to correspond
to a simple pole of the exact propagator. Thus,
$S^{-1}(\omega,{\bf p})$ vanishes on shell, and the second
term in eq.~(\ref{DS2}) gives no contribution.
The leading contribution to the on-shell self-energy reads then
\beq\label{DS3}
 \Sigma({\rm on-shell})\,\simeq\,-g^2T 
 \int \frac{{\rm d}^3q}{(2\pi)^3}\,\frac{1}{q^2 + \mu^2}\, 
\frac{1}{{\bf v}\cdot ({\bf q}) - i\eta}\,\simeq\,
i\,\frac{g^2T}{4\pi}\,\ln \frac{\omega_p}{\mu}\,,\eeq
and coincides with the IR singular part
of the one-loop self-energy. This is only possible 
if the aforementioned compensation of the leading power-like
divergences holds in all orders. Note that the above arguments
become meaningless in the physical limit $\mu\to 0$,
where not only does the estimate (\ref{DS3}) become
logarithmically divergent, but the integral multiplying
$S^{-1}(\omega,{\bf p})$ also  diverges on the mass-shell.

\end{document}